%% file: ms.tex
\documentclass[12pt, preprint]{aastex}

\newcommand{\lsim}{\ \raise -2.truept\hbox{\rlap{\hbox{$\sim$}}\raise
5.truept\hbox{$<$}\ }}
\newcommand{\gsim}{\ \raise -2.truept\hbox{\rlap{\hbox{$\sim$}}\raise
5.truept\hbox{$>$}\ }} \newcommand{\grv}{\`}

\lefthead{Surface Brightness Fluctuations} \righthead{}
\shortauthors{Biscardi et al.}

\begin{document}

\title{Optical Surface Brightness Fluctuations of shell galaxies
towards 100 Mpc}

\author{ Biscardi I. \altaffilmark{1,2}, Raimondo G.\altaffilmark{1},
Cantiello, M.\altaffilmark{1,3}, Brocato, E.\altaffilmark{1}}
\altaffiltext{1}{INAF-Osservatorio Astronomico di Teramo, Via M.
Maggini s.n.c., I-64100 Teramo, Italy; biscardi, raimondo, cantiello,
brocato@oa-teramo.inaf.it } \altaffiltext{2}{Dipartimento di Fisica -
Universit\grv a di Roma Tor Vergata, via della Ricerca Scientifica 1, 00133
Rome, Italy .}  \altaffiltext{3}{Department of Physics and Astronomy,
Washington State University, Pullman, WA 99164.}

\begin{abstract}

We measure $F814W$ Surface Brightness Fluctuations (SBF) for a sample
of distant shell galaxies with radial velocities ranging from 4000 to
8000 km/s. The distance at galaxies is then evaluated by using the SBF
method.  For this purpose, theoretical SBF magnitudes for the ACS@HST
filters are computed for single burst stellar populations covering a
wide range of ages (t=1.5-14 Gyr) and metallicities (Z=0.008-0.04).
Using these stellar population models we provide the first
$\bar{M}_{F814W}$ versus $(F475W-F814W)_0$ calibration and we extend
the previous I-band versus $(B-I)_0$ color relation to colors
$(B-I)_{0}\leq 2.0$ mag.  Coupling our SBF measurements with the
theoretical calibration we derive distances with a statistical
uncertainty of $\sim 8\%$, and systematic error of $\sim 6 \%$.  The
procedure developed to analyze data ensures that the indetermination
due to possible unmasked residual shells is well below $\sim 12
\%$. The results suggest that \emph{optical} SBFs can be measured at
$d \geq 100 Mpc$ with ACS@HST imaging. SBF-based distances coupled
with recession velocities corrected for peculiar motion, allow us
obtain $H_{0} = 76 \pm 6$ (statistical) $\pm 5$ (systematic)
km/s/Mpc.

\end{abstract}

\keywords{galaxies: elliptical and lenticular, cD --- galaxies:
distances --- galaxies: photometry --- galaxies: fundamental parameters --- cosmology: distance scale}

\section{Introduction}
\label{sect: intro}

The Surface Brightness Fluctuation (SBF) method is a powerful technique
to derive distance to galaxies when individual stars cannot be
resolved. The method was first introduced by \citet{TonrySchneider88}
and is based on a simple concept: the poissonian distribution of
unresolved stars in a galaxy produces brightness fluctuations between
pixels of the galaxy image. The resulting pixel-to-pixel variance of
the fluctuation is inversely proportional to the square of the
distance, and therefore can be used as distance indicator. Although
limited by pixel resolution and high S/N ratio, the SBF method showed
the capability of spanning an interval of distances extremely wide,
ranging from Local Group objects
\citep{Ajhar+94,Raimondo+05b,Rekola+05} up to galaxies at
100 Mpc \citep{Thomsen+97,Jensen+01}.

Ideally, SBF measurements can be derived for almost any morphological
type of galaxy, provided that the region analyzed has a nearly
regular/smooth luminosity profile.  Indeed, the presence of
morphological irregularities represents a constrain to reliable
measurements of SBF, for example, if dusty regions are present, the
application of SBF technique requires an accurate masking of dust.

As recognized by \citet{Tonry+90}, the properties of the stellar
populations in the galaxy cannot be neglected in calibrating the
absolute SBF magnitude. This is usually overcome by providing a
relation between the SBF magnitude and a galaxy color. Once such a
calibration is available, the SBF method can be adopted to infer the
distance of elliptical, lenticular and spiral galaxies with prominent
bulge \citep[e.g.][]{Tonry+01,Mieske+06,Mei+07}. Moreover, the SBF
technique has been successfully applied to low surface brightness
dwarf ellipticals \citep{Jerjen+98,Jerjen+00a,Jerjen+00b, Rekola+05},
and Galactic or Magellanic Clouds stellar clusters \citep{Ajhar+94,
Gonzalez+04, Raimondo+05b}.

An interesting case is represented by shell elliptical galaxies, a
small number of which is included in the sample observed by
\citet{Tonry+01}.
Shell structures are generally faint and sharp stellar features,
and they are considered a robust indicator of past merging or
interaction events \citep{Malin&Carter83, Wilkinson+00}. The stellar
population in the shell depends on the galaxy with which the merger
has taken place and on the time spent since the shell structure has
formed \citep{Pence86, Wilkinson+00}. Many authors have found
shells redder or bluer than the underlying galaxy
\citep{Forbes+1995, Turnbull+99, Sikkema+07}.

It is reasonable to expect that the presence of shells might influence
the SBF signal of the galaxy. In this respect, the high quality of
Advanced Camera for Surveys (ACS) images is crucial to remove the
shell structure, allowing the measurement of SBF of the galaxy, even
at high distances. Thus, the SBF measurement remains a relevant tool
to investigate the distance of shell galaxies.

In the present work, we derive SBF magnitudes of a sample of distant
shell-ellipticals taking advantage of the high resolution power of the
ACS on board of the Hubble Space Telescope (HST). We select deep ACS
images and measure F814W-band ($\sim $ I-band) SBF of the four
galaxies: PGC\,6510, PGC\,10922, PGC\,42871 and PGC\,6240, with
recession velocities reaching $\approx$ 8000 km/s. The main goal of
our work is to measure their distance by using the SBF technique. No
previous distance determination is available for any of these
galaxies, except for the kinematic distance modulus. The farthest
object of the sample, at a distance of $\sim$ 100 Mpc, lies near the
upper limit of distances obtained using optical SBF measurements
\citep{Thomsen+97,Mei+03}. It is worth noticing that none of these
objects has been previously analyzed in detail, with the only
exception of the recent work of \citet{Maybhate+07} on the globular
cluster system of PGC\,6240.

In the next section we present the observational data and the general
properties of the selected galaxies. The procedure to derive the SBF
measurements and the data analysis are described in Sect.  \ref{sect:
dataanalysis}. The results are presented and discussed in
Sect. \ref{sect: discussion}, where we provide new SBF predictions
together with new calibrations of absolute SBF magnitudes for the
ACS@HST filters based on single-burst stellar population (SSP)
models. In the same section, an evaluation of $H_0$ is also
obtained. A brief summary  is given in Sect. \ref{sect:
conclusions}.

\section{The observational data}
\label{sect: data}

The HST images used in the present work were obtained with ACS in
its Wide Field Channel mode. The large field of view, high resolution
and sharp point-spread function (PSF) characterizing ACS images are
critical requirements to attempt the SBF measurements of the distant
galaxies we have selected.

The images were retrieved from HST archive. A requirement in selecting
the data is that the SBF S/N ratio in the F814W band is $\gsim$ 5,
as suggested by \citet{Blakeslee+99}. This condition is verified for
the three ellipticals PGC\,6510, PGC\,10922, PGC\,42871, while
PGC\,6240 images have slightly lower S/N ($\approx$4). For these
galaxies F475W images were also available, though the  exposure times
prevented their use to measure SBF (S/N $<<$ 5).  However,
they were retrieved and analyzed to obtain integrated magnitudes
and colors. The  images are associated with proposal GO
10227 (PI: P. Goudfrooij) designed to study the globular cluster
system of the four giants, post-starburst shell ellipticals. The main
properties of the selected galaxies are reported in Table
\ref{table:obs}.

We downloaded the ACS images processed with the standard calibration
pipeline (CALACS), that includes bias, dark and flat-fielding
corrections. The images from the archive still require the final image
combination and the correction to remove the geometric distortions. To
this purpose we used the PYRAF task $\textit{multidrizzle}$
\citep{Koekemoer+02}, that also provides the bad pixel identification
and the cosmic-ray rejection.  No sky subtraction was performed at
this stage. To reduce the effect of noise correlation introduced by
the drizzling procedure \citep{Fruchter&Hook02}, we adopted the
LANCZOS3 kernel which, as demonstrated by \citet{Mei+05a} and
\citet{Cantiello+05}, is adequate for the purpose of SBF estimation.

\section{Data Analysis}
\label{sect: dataanalysis}

To derive the SBF measurements we follow the same basic procedure
adopted in \citet{Cantiello+05,Cantiello+07a}. In this section, we
summarize the relevant steps  to measure SBF magnitudes and we
enlighten the differences from the quoted works.

First, a provisional sky value is derived from the corner with lowest
sky counts, and a mask of the bright sources (saturated stars,
extended galaxies, etc.) is obtained. Then, the brightness profile of
the galaxy is modeled using the IRAF\footnote{IRAF is distributed by
the National Optical Astronomy Observatories, which are operated by
the Association of Universities for Research in Astronomy, Inc., under
cooperative agreement with the National Science Foundation.}/STSDAS
task ELLIPSE \citep{Jedrzejewski+87}. After the galaxy model is
subtracted, we derive a new mask of the faint sources which clearly
appears after the galaxy light is removed. This new mask is fed to
ELLIPSE to obtain a new galaxy model.

We use the isophotal geometry, as derived with the last galaxy
model, to get the surface brightness and color profiles of the
galaxy. The galaxy profile is fit using a de Vaucouleurs law to find
the sky as the zeropoint constant in the fit. For all galaxies, we
find that a de Vaucouleurs law is well suited in the 1-5$\arcsec$
region of the galaxy. This result confirms the known evidence that
the photometry of shell galaxies is the same expected for a normal
elliptical \citep[e.g.][]{Wilkinson+00}.

\begin{center}
\begin{figure*}[t]
\epsscale{.8}\plotone{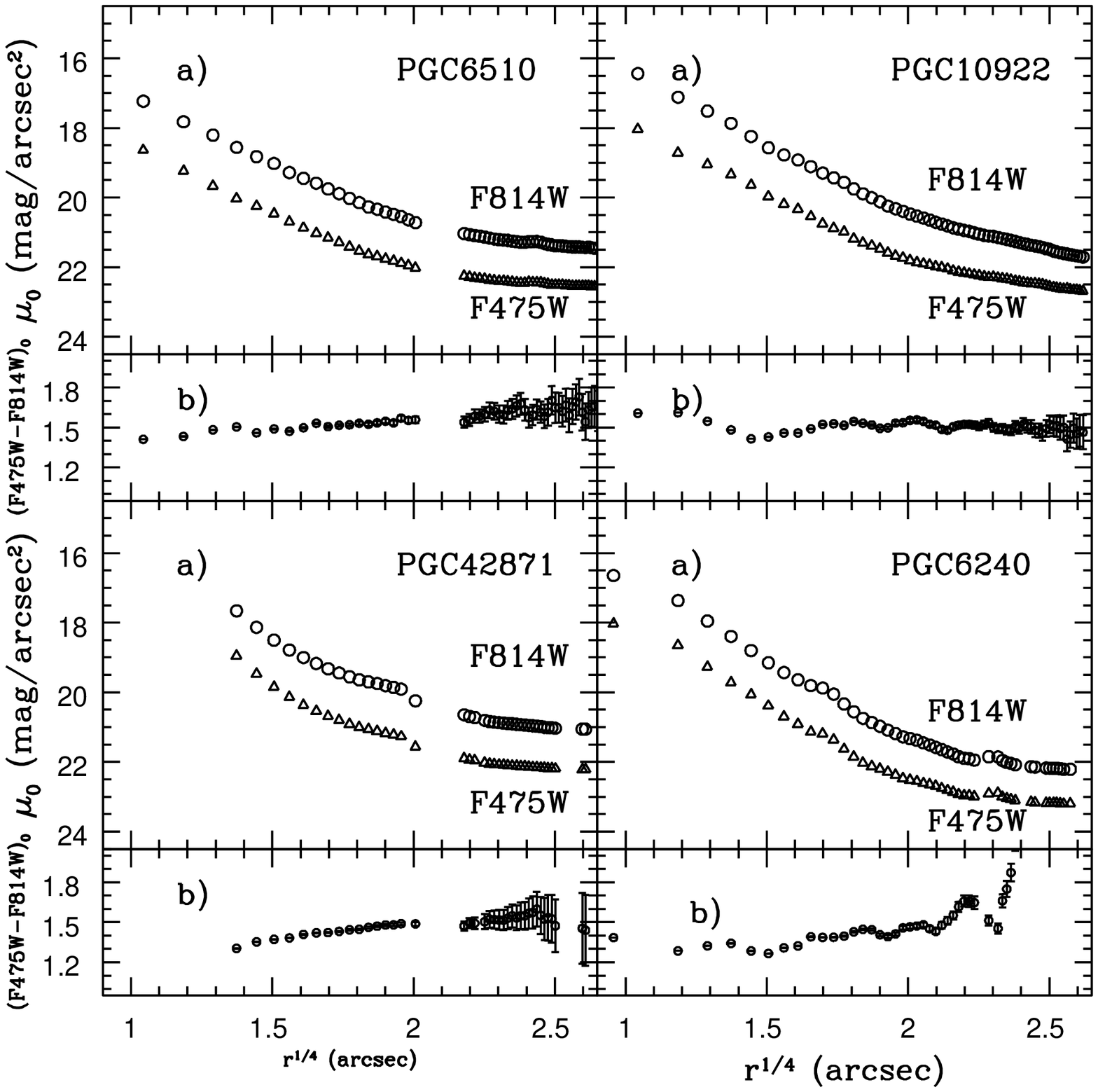} \caption{ Panels $a)$: F814W/F475W
band surface brightness profile as function of $r^{1/4}$ (open
circles/triangles). Panels $b)$: measured (B-I)$_{0}$ profile as
function of $r^{1/4}$. Data are corrected for Galactic extinction and
K-correction.}
\label{fig:mu}
\end{figure*}
\end{center}

The new sky value is adopted and all the above steps are iterated,
till convergence. At the end of these iterations we have the final sky
value, the mask of external sources, and the best galaxy model. The
final surface brightness and color profiles for the four galaxies are
shown in Fig. \ref{fig:mu}. It is worth noting that  the average
color of the four galaxies is bluer than in normal ellipticals, and
 the color gradient is positive or nearly flat. Both
characteristics can be considered normal for this class of galaxies
\citep{Tamura+00,Lee+06}. For all data we apply a K-correction
$K_{F814W}\approx K_{I} \approx 0.5 \cdot z$ and $K_{F475W} \approx
K_{g}\approx 2 \cdot z$ \citep{Poggianti97} for integrated color. The
extinction correction are evaluated using the prescriptions given by
\citet[][their Table 14]{Sirianni+05}.

We subtract the sky value and the galaxy model from the original
image. This operation leaves a large scale residual background, due to
mismatch of the real galaxy with the model.  The large scale residuals
are removed using the background map derived with the photometry
package SExtractor \citep{Bertin+96}. We have carried out some
numerical experiments to determine the best background map parameters
able to provide both a good subtraction of the large scale residuals,
and the best possible removal of the shell features. After several
experiments we have chosen a mesh size of 15$\times$15 pixels$^2$
(BACK\_SIZE=15), with 3 background-filtering meshes
(BACK\_FILTERSIZE=3).

Up to this point, the procedure is applied to both filters.  The sky,
galaxy model and large scale residuals are subtracted to the original
frame to derive the {\it residual} frame.  Note that possible dust
regions, as found in PGC\,42871, are masked
out since they could compromise the SBF measurement. Such regions are
better recognized from the F475W image, the same dust mask is used for
both filters.

We run SExtractor on the residual frame to obtain the photometric
catalog of external sources (foreground stars, globular clusters and
background galaxies).  This photometric catalog is used to construct
the luminosity functions (LFs), which will be used to estimate the
contribution to the fluctuations coming from faint unmasked external
sources ($P_{res}$).

\begin{center}
\begin{figure*}[t]
\epsscale{.8}\plotone{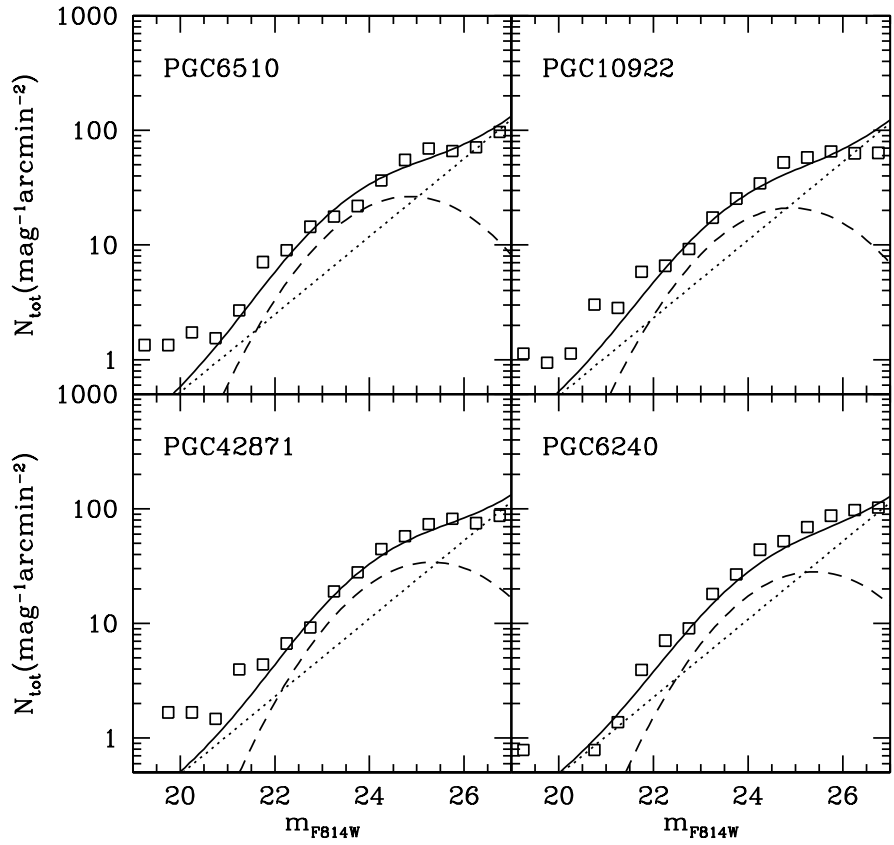} \caption {The LFs of external
sources. Open squares mark observational data; model fits to globular
cluster and galaxy LFs are shown as dashed and dotted lines,
respectively, and their sum as solid line.} \label{fig:LF}
\end{figure*}
\end{center}

 The procedures used to fit the LFs are the same described in
\citet[][]{Cantiello+05}. In particular, we adopted $M_I = -8.5$ mag
as the absolute Turn Over Magnitude (TOM) of the GCLF
\citep{Harris+01}, while the exponent for the power law LF of
background galaxies is $\gamma=0.34$ \citep{Bernstein+02}. These
parameters are used to start an iterative fitting process where the
surface density of galaxies and globular clusters, and the galaxy
distance are allowed to vary until their best values are found via a
maximum likelihood method. Fig. \ref{fig:LF} shows the best fit of
the observed LFs obtained for the four galaxies,  used to derive
$P_{res}$ .

 Before computing the power spectra of the residual image and of
the PSF, and evaluate their azimuthal averages, we mask regions with
residual contamination from shells, and divide the residual frame by
the square root of the galaxy model. It is worth noting here that only
the shells not completely removed by the background map subtraction,
i.e. the most prominent shells, required further masking.

\begin{center}
\begin{figure*}[t]
\epsscale{.8}\plotone{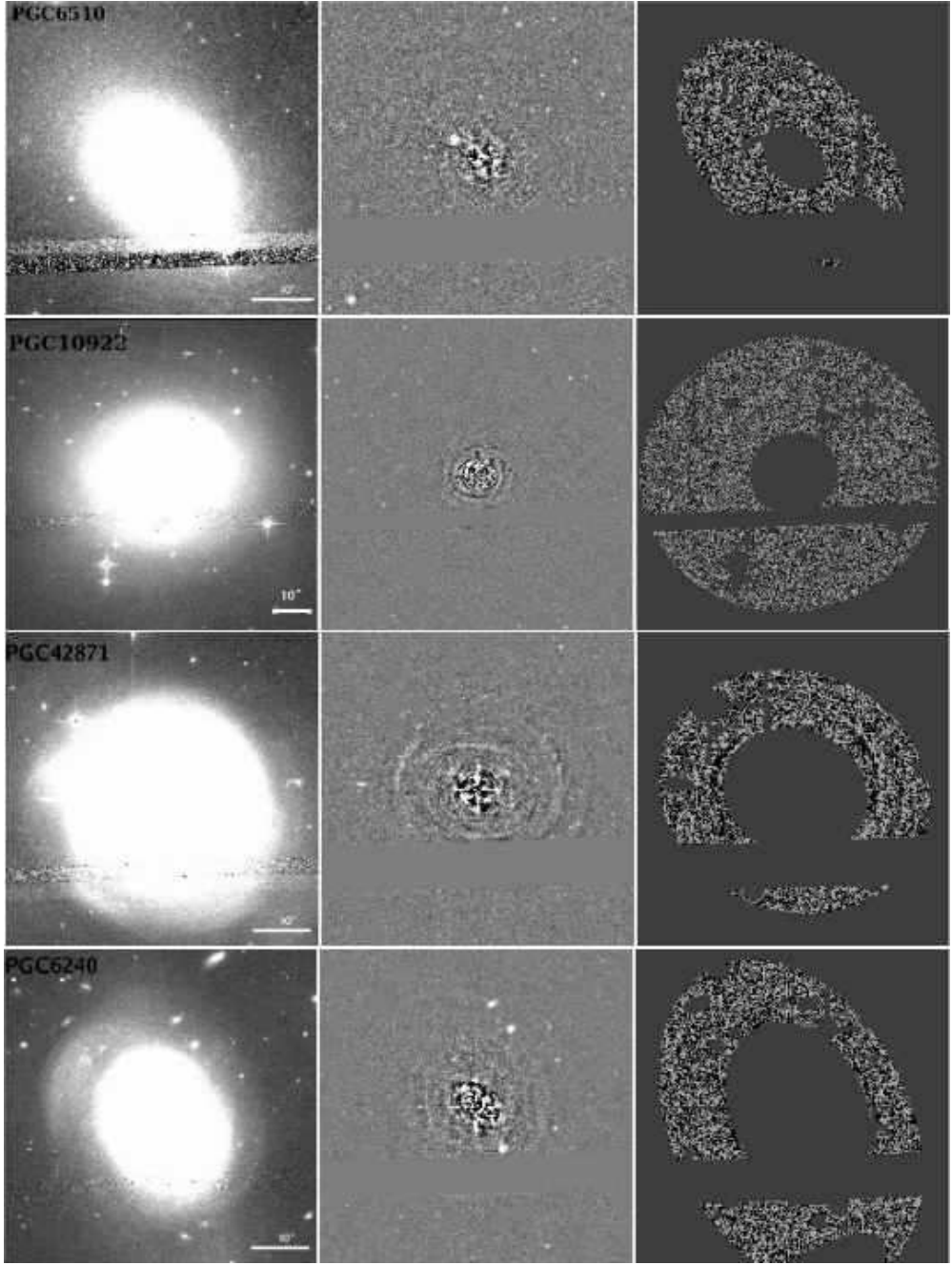}
 \caption { From left to right: the
original I-band frame, the residual frame and residuals times the
final adopted mask, for each galaxy (upper left quote in each
panel).}
 \label{fig:images}
\end{figure*}
\end{center}

The frames obtained with the analysis described up to this point are
shown in Fig. \ref{fig:images}.  We point out that in general the
residuals appear similar to typical residual frames of SBF
measurements of normal ellipticals.  Nevertheless, in Fig.
\ref{fig:images} we adopted a scale to emphasize the regions where the
procedure described failed in the complete subtraction of shells (in
particular, PGC42871 and PGC6240).  To quantify the additional
systematic uncertainty due to unsubtracted shells, we performed a
specific test described at the end of this section.

 Next, the total fluctuation amplitude is determined as the
constant factor $P_{0}$ multiplied by the PSF power spectrum, to match
the power spectrum of the residual frame P(k):
\begin{equation}\label{eq:pk}
P(k)=P_{0}\cdot E(k) \ + \ P_{1},
\end{equation}
where $P_{1}$ is the white-noise component, and $E(k)$ is the
convolution between the PSF power spectrum and the power spectrum of
the mask function. The latter mask also takes into account the shape
of the galaxy annulus adopted for the SBF measurement. The minimum
annular radius has been fixed to be the minimum radius without dust
contamination, while the maximum radius is fixed to the region where
the galaxy to sky counts ratio is $\gsim$ 1\footnote{ As
\citet{Cantiello+07a}, we consider one single annulus per galaxy,
because of the lower S/N ratio of the images, and because of the, on
average, small available area. It must be noted that in our procedure
the contribution of the external sources is evaluated taking into
account their radial position, so that using one single annulus does
not introduce any systematics on $P_{res}$. Moreover, we have carried
out a three-annuli measurement on the galaxy with the largest spatial
extension, PGC\,6510, as a result we found that the averaged value
agrees within uncertainties with the measurement reported in Table
\ref{tab:rismag}.}.

A robust linear least-squares method \citep{Press+92} is used to
fit Eq. (\ref{eq:pk}). The lowest k-numbers ($k < 250$), that have
been corrupted by the subtraction of the smooth background profile
\citep{Blakeslee+99}, and the high k-numbers ($k> 600$), that have
been corrupted by the drizzling procedure, are excluded from the
fitting  \citep[see][for more details]{Cantiello+05,Cantiello+07a}.

\begin{center}
\begin{figure*}[t]
\epsscale{.8}\plotone{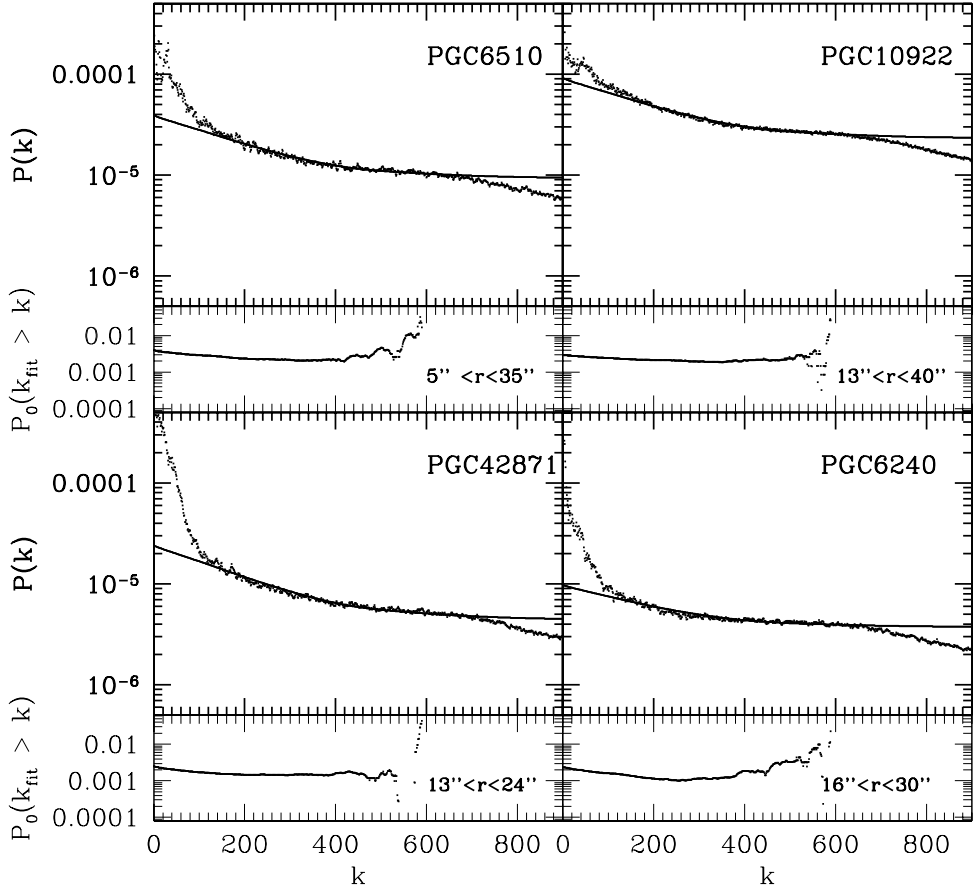} \caption{For each labeled galaxy,
the top panels show the azimuthal average of the power spectrum
(the average radii of the annuli used are reported in each panel).
The observational data (dots) are fitted by the sum of a scaled
PSF power spectrum plus the constant white noise term (solid
line). The lower panels show the $P_{0}$ obtained as a function of
the starting wavenumber of the fit, $P_{0}(k_{fit}>k)$. The final
$P_{0}$ adopted is the weighted average of values in the flatter
$P_{0}$ vs $k$ region. We excluded the lowest $k$-numbers, that
have been corrupted by the background subtraction, and the highest
$k$-numbers, that have been corrupted by the drizzling procedure.}
\label{fig:ps}
\end{figure*}
\end{center}

Fig.  \ref{fig:ps} exhibits the azimuthal average of the power
spectrum for each galaxy, the best fit lines are also shown.

Finally, the SBF magnitude is evaluated as:
\begin{equation}\label{eq:mbar}
\overline{m}_{F814W}=-2.5 \log \left(P_{0}-
P_{res}\right)+m^{\ast}-A_{F814W}-K_{\overline{F814W}} \nonumber
\end{equation}
where $P_{res}$ is the extra contribution of unmasked external
sources, evaluated from the fitted LF as described in
the quoted papers. The amplitude of
$P_{res}$ is small for all galaxies of the sample, being on average
 $P_{res}/P_0 \sim 0.08$ (Table \ref{tab:rismag}). $m^{\ast}$ is
the zero-point ACS magnitude in the VEGAMAG system reported by
\citet[][$m^{\ast}_{F814W}=25.501 \ mag $]{Sirianni+05}, $A_{F814W}$
is the extinction correction in the F814W passband, and
$K_{\overline{F814W}}$ is the $K$-correction term. We apply
$K_{\overline{F814W}}\approx 7 \cdot z$ after \citet{Thomsen+97}.

We used \citet{Sirianni+05} equations to transform the F814W and F475W
into the standard B and I magnitudes. However, in the forthcoming
section we will also take into account the magnitudes in the ACS
photometric system.

The main differences between the present data and the images used by
\citet{Cantiello+05,Cantiello+07a} are $i)$ the much greater distance
of the objects, which affects the LF fitting, and $ii)$ the rather
complex shell structure. For this reason we have performed two
additional tests with respect to the ones described in the quoted
papers.

 The first one is related to the Globular Cluster LF (GCLF).  The
data quality of the images in some cases does not allow to reliably
sample the TOM of the GCLF. This is confirmed by the recent work
published by \citet{Maybhate+07} on the GCLF of PGC\,6240, based on
the same data used here. These authors do not provide an estimation of
the TOM, however, from a visual inspection of their Fig. 11, the
extrapolated TOM appears $\approx$1 magnitudes fainter than the one
adopted here. Since the large uncertainty, we perform a specific test
to evaluate its effect on the SBF. Adopting the GCLF of
\citeauthor{Maybhate+07}, the SBF magnitude changes a few
hundredths of mag ($\lsim 0.05$ mag),  that is a $\sim$ 3\%
uncertainty on the distance.

The second test concerns how the presence of shells affects
SBF. To evaluate this effect we selected the farthest galaxy in the
sample (PGC6240), then we run the whole SBF measurement procedure on
the image, with and without masking the most prominent shell. The
result of this test shows that the final SBF magnitude changes by
$\lsim$0.25 mag. Even though this is not an exhaustive evaluation of
the uncertainty caused by shell features, it is a robust suggestion
that the adopted procedure should keep this source of uncertainty
fairly below 0.25 mag. Thus, we consider this value as the
upper limit of systematic uncertainties due to possible unmasked
residual shells.

In conclusion, the total statistical uncertainty on the SBF is
obtained as the square sum of the fitting uncertainty on $P_{0}$, and
the default 25\% uncertainty on $P_{res}$ \citep{Tonry+90}. The effect
of the sky uncertainty is negligible on the SBF, while it is the main
source of error in the color. In addition, these measurements suffer
for a total systematic uncertainty due to: $1)$ the PSF normalization
($\approx$ 0.03 mag), $2)$ the fit of LFs ($\approx$ 0.05 mag), $3)$
the filter zero points ($\approx$ 0.01 mag), and $4)$ if necessary,
the transformation from the ACS photometric system to the standard
system ($\approx$ 0.02 mag). Summing in quadrature all the systematic
sources of errors, we find that a total systematic uncertainty
$\approx 0.1$ mag affects our SBF measurements \citep{Cantiello+05}.
On average, this corresponds to a systematic error of $\approx 6\%$ on
distances and on the Hubble constant. If we also add the maximum
systematic uncertainty of $\approx0.25$ mag due to the presence of
shells, the total systematic error becomes $\approx 12\%$. It is worth
emphasizing here, again, that the 0.25 mag systematic uncertainty due
to the shells is an upper limit, as it has been derived from the worst
case, i.e. farthest distant galaxy, prominent shells.

Table \ref{tab:rismag} reports the final SBF measurements and
 statistical uncertainties for all galaxies. In the Table
we also report the SBF and colors obtained transforming the ACS
filters F475W and F814W into standard B and I passbands.

\section{Discussion}
\label{sect: discussion}

\subsection{Calibration, SBF models and distances}
\label{subsect: calibration}

The application of SBF method as distance indicator requires the
calibration of the absolute SBF magnitude, $\overline{M}$, versus the
broadband color. To date, no calibration is available for the ACS
F814W bandpass versus $(F475W-F814W)_{0}$ color, in the color range of
the present galaxies sample.  Moreover, even transforming the
$(F475W-F814W)_{0}$ color to the standard $(B-I)_{0}$ color, the
empirical calibration of SBF in the standard I band determined by
\citet{Cantiello+05} is defined in the color range $2.0 \leq
(B-I)_{0}\leq2.25$, while all our galaxies have $(B-I)_{0}\leq 2.0$
mag.  Thus, we decided to extend to bluer colors the calibration of
absolute SBF magnitudes, by using models.

The SBF models used here are based on the most updated version of the
code SPoT \citep[Stellar Population
Tools,][]{Raimondo+05b}\footnote{http://www.oa-teramo.inaf.it/SPoT}.
These models have the advantage of fitting the SBF and color of
ellipticals, as well as of resolved and unresolved stellar clusters,
for a wide range of ages and metallicities (see
\citeauthor{Raimondo+05b} and references therein for details). For the
specific purpose of this study, we computed the theoretical SBFs in
the ACS VEGAMAG photometric system using the BaSeL 3.1
\citep{Westera+02,Patricelli+06} stellar atmospheres library. The resulting SBF
magnitudes are reported in Table \ref{tab:models}.

\begin{center}
\begin{figure*}[t]
\epsscale{1.}\plottwo{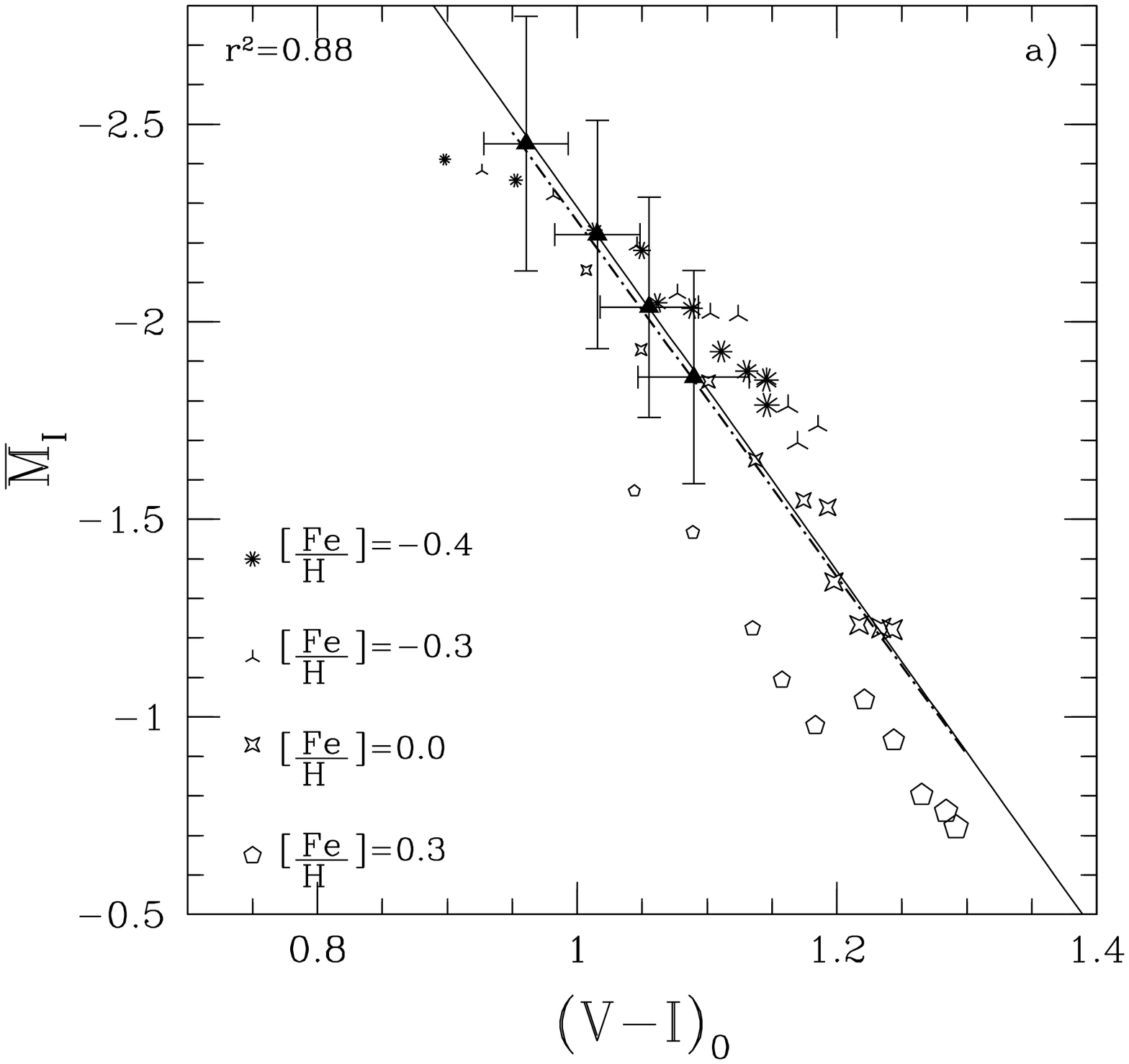}{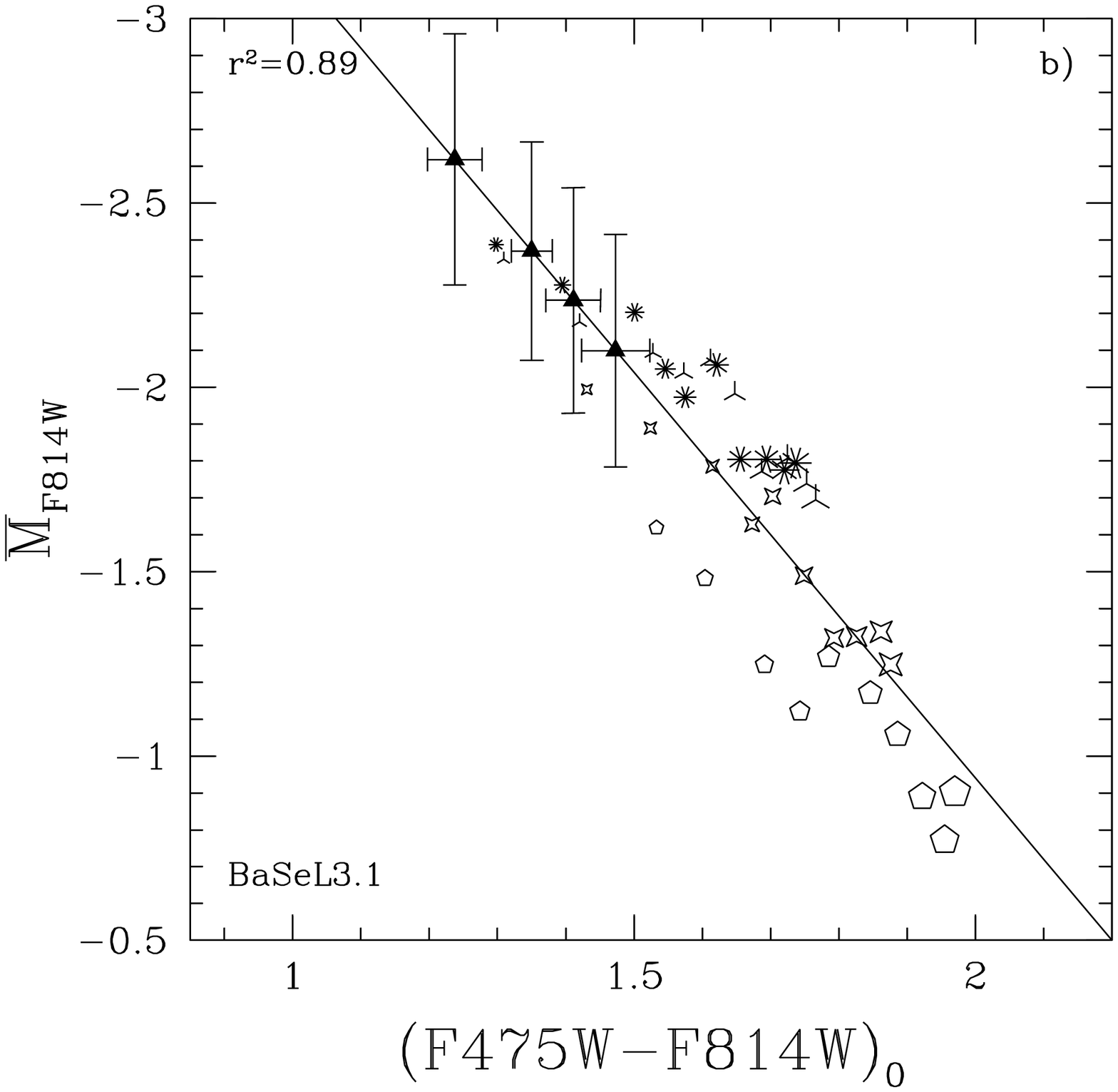}
\epsscale{2.2}\plottwo{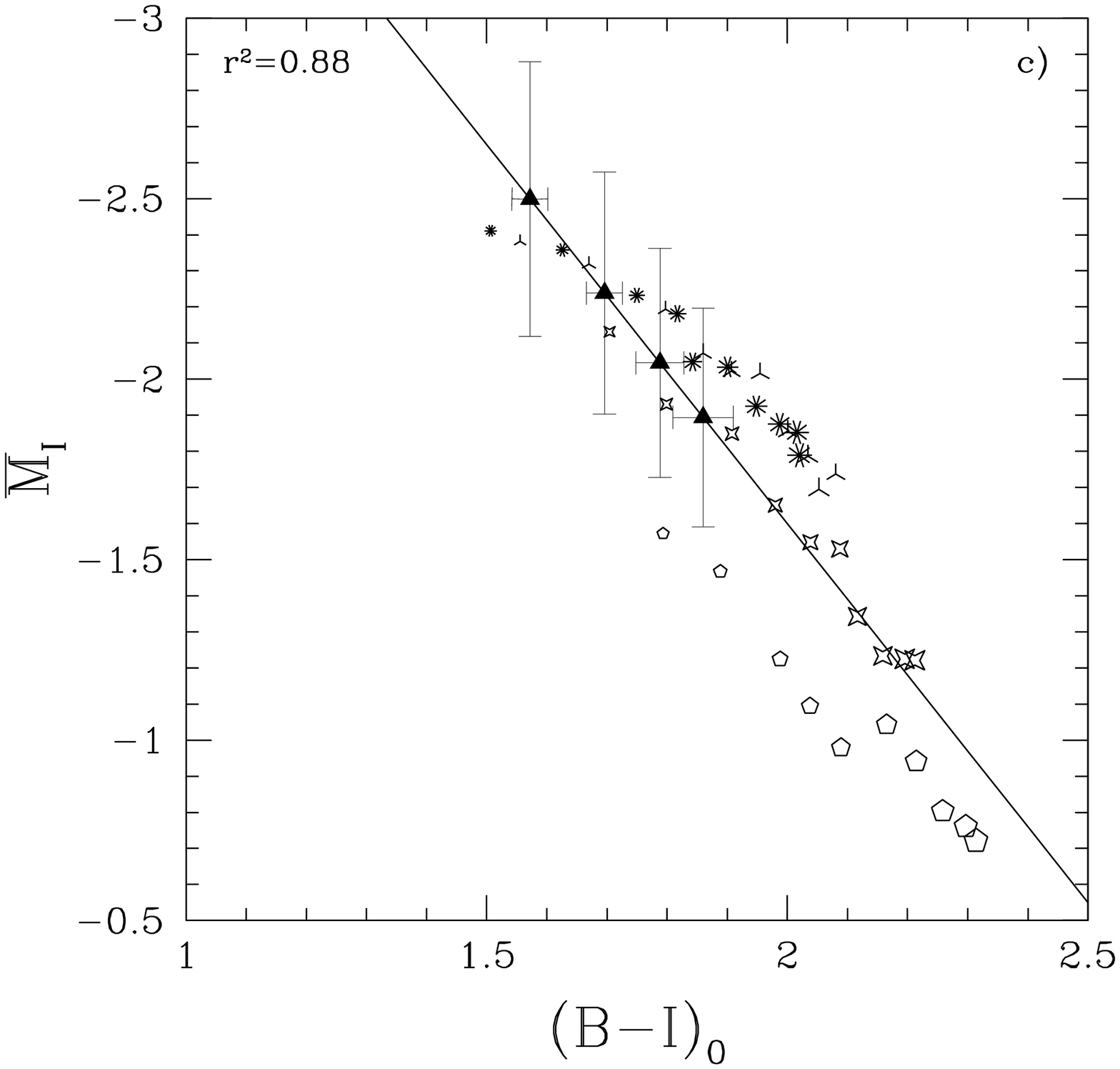}{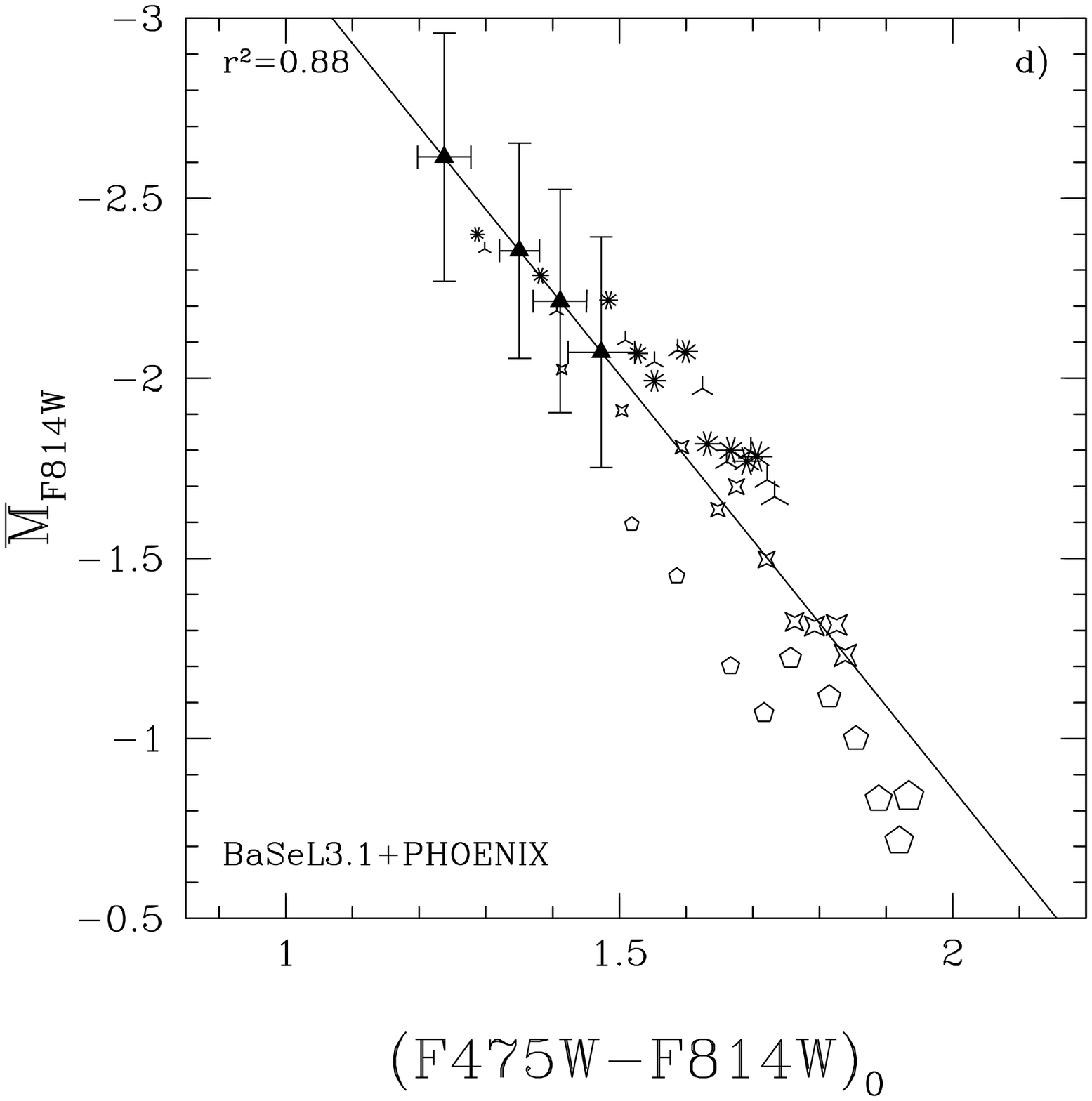}
\caption {Absolute SBF magnitudes versus integrated colors. In the
figure we plot our  models  of different
metallicities (as labeled) and ages (t=1.5, 2, 3, 4, 5, 7, 9, 11, 13,
and 14 Gyr). Symbols with increasing size mark models of older
age. Full triangles mark the observational data. In each panel we
report the linear fit to models (solid line) and the correlation
coefficient $r^{2}$ (upper left label). \textit{Panel a:} The
calibration of $\overline{M}_{I} $ versus $(V-I)_{0}$ from
\citet{Jensen+03} is shown with a dot-dashed line. \textit{Panel b:}
As in panel ($a$) but the absolute $\overline{M}_{F814W} $ magnitude
and $(F475W-F814W)_{0}$ color are used. \textit{Panel c:} As in panel
($a$) but the $(B-I)_{0}$ color is used. \textit{Panel d:} As in panel
($b$) but a different stellar atmosphere library is used (upper right
label).}
\label{fig:fit}
\end{figure*}
\end{center}

We start with the empirical calibration by \citet{Jensen+03}:

\begin{equation}
\label{eq:jensen} \overline{M}_{I} = (-1.58 \pm 0.08) + (4.5 \pm
0.25) \times [(V-I)_{0}-1.15]
\end{equation}

\noindent defined in the color range $0.95 \leq (V-I)_{0} \leq 1.3$.
This relation has a high degree of reliability, since the zeropoint
magnitude has been obtained using the improved period-luminosity
relations for Cepheids by \citet{Udalsky+99}, and the slope is the one
derived by \citet{Tonry+97} based on group membership of galaxies. On
the theoretical side, we fit the SBF versus color models with a
 robust straight line fitting process that minimizes the mean
absolute deviation. We selected stellar population models in the age
range $1.5 \leq t(Gyr) \leq 14$, and the metallicity range $-0.4 \leq
\left[Fe/H \right] \leq 0.3$ (Fig. \ref{fig:fit}, panel $a$). As a
result we obtain:

\begin{equation}
\label{eq:myjensen} \overline{M}_{I}= (-1.6 \pm 0.1)+ (4.5 \pm
0.2)\times [(V-I)_{0}-1.15]
\end{equation}
in very good agreement with Eq. \ref{eq:jensen}.

\begin{center}
\begin{figure*}[t]
\epsscale{.9}\plotone{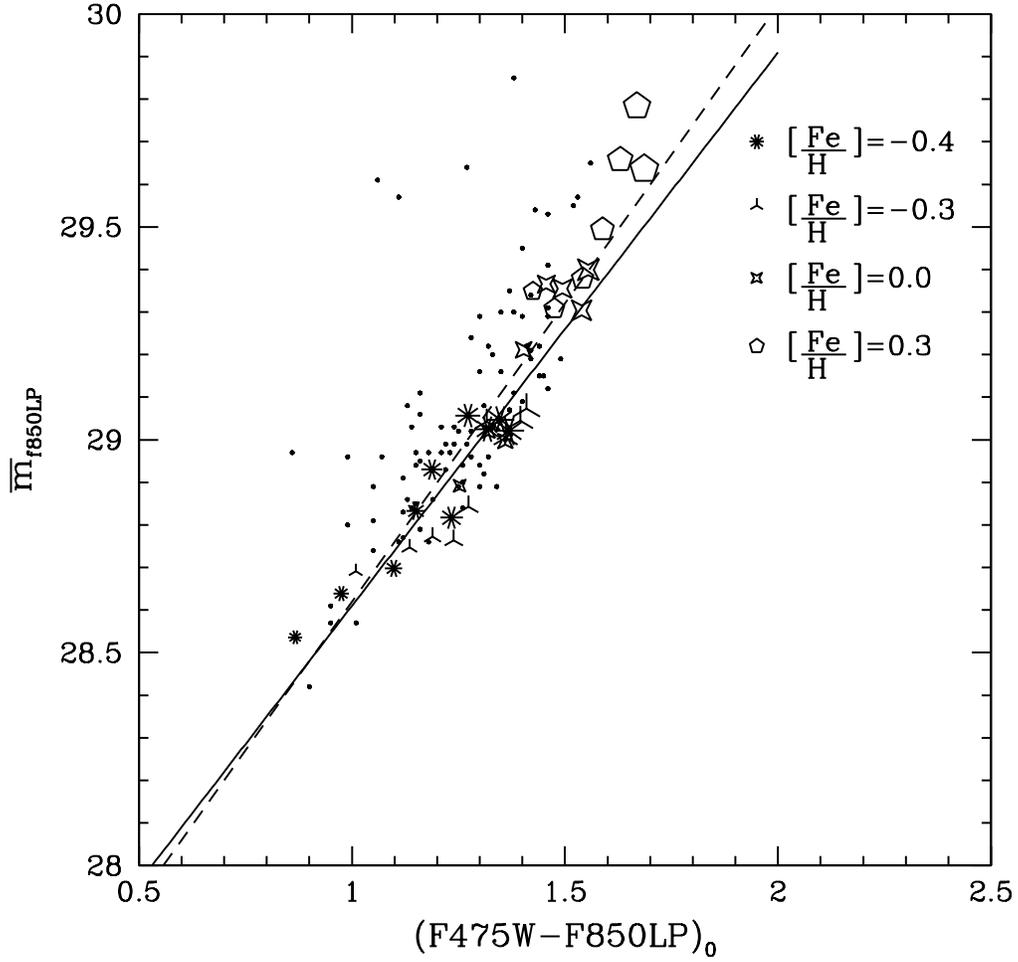}
\caption { $\overline{m}_{F850LP}$ versus $(F475W-F850LP)_{0}$. In
the figure we plot the new SBF models in the ACS photometric system
(ABMAG), shifted to a Virgo distance modulus of 31.1(symbols are
as in Fig. \ref{fig:fit}). The observational data (full dots) are
from \citet{Mei+07}. The best fit of the present SBF models is shown
as a solid line, the best fit of the data by \citet{Mei+07} is
reported as a dashed line.}
\label{fig:CFRmei}
\end{figure*}
\end{center}

 The same models, but for the ACS photometric system, are compared
to the observational data from the ACS Virgo Cluster Survey
\citep{Mei+07}. We have adopted a DM$\simeq$31.1 for the Virgo cluster
\citep{Ferrarese+00}, and transformed models into the ABMAG
photometric system for sake of homogeneity with data (Fig.
\ref{fig:CFRmei}). A linear fit to models provides a slope $\alpha=
1.4 \pm 0.1$ and intercept $\beta=29.0 \pm 0.1$, in good agreement
with the empirical fits provided by Mei and collaborators, who give
$\alpha=1.3 \pm 0.1$ and $\beta=29.09 \pm 0.04$. We have also checked
the case of a double linear fit, as suggested by the authors. We
obtained $\alpha=0.8 \pm 0.1$ and $\beta=29.0\pm 0.1$ in the color
interval $1.0\leq (F475W-F850LP)_0 \leq 1.3$; while $\alpha=1.7\pm0.1$
and $\beta=29.0\pm 0.1$ in the color range $1.3< (F475W-F850LP)_0 \leq
1.6$. In both cases the fit from models agrees within $1\sigma$ with
the empirical calibrations.

\begin{center}
\begin{figure*}[t]
\epsscale{.9}\plotone{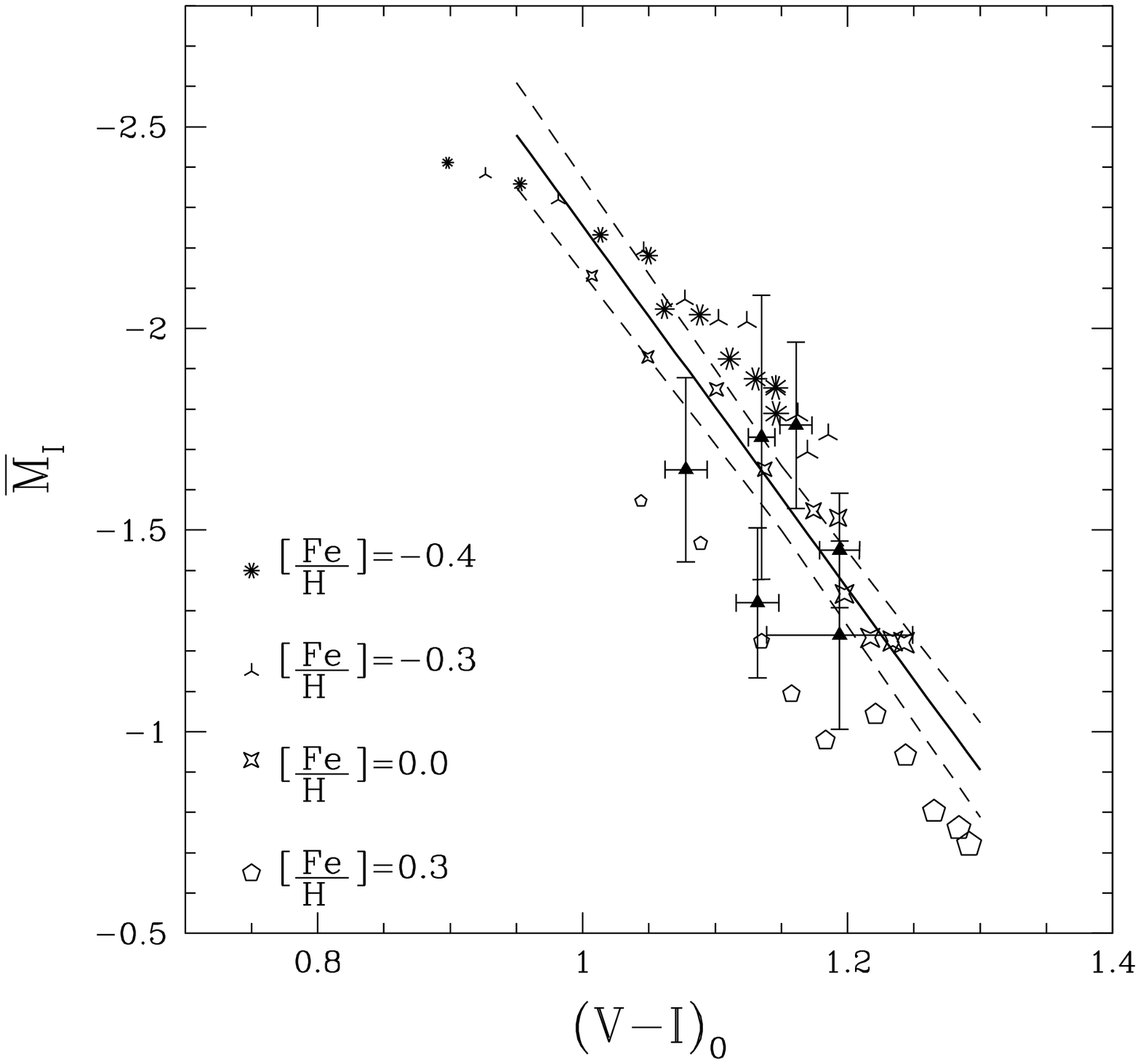}
\caption {$\overline{M}_{I} $ versus $(V-I)_{0}$. Symbols are as in
 Fig. \ref{fig:fit}. In the figure we plot the calibration from
 \citet[][solid line]{Jensen+03}, 1$\sigma$ dispersion lines are
 reported with  dashed line. A sample of shell elliptical
 galaxies from \citet{Tonry+01} are reported as triangles. The
 absolute SBF magnitudes are derived coupling the apparent SBF
 measurements, with the SBF-independent DMs reported in Table
 \ref{table:shell}. It is noticeable the absence of any systematic
 deviation of shell ellipticals with respect to the empirical
 calibration ($\langle \bar{M}_{J03}-\bar{M}_{NON-SBF~DMs}\rangle =
 -0.06 \pm 0.22$), and to the stellar population models.}
\label{fig_shell:ps}
\end{figure*}
\end{center}

Before going further, we evaluate the capability of the SBF method to
derive accurate distances of shell ellipticals, as their peculiar
morphology might disturb the SBF measurement. To this purpose, we
selected a sample of shell galaxies from the \citet{Tonry+01}
database, for which the distance modulus (DM) is available from
methods not related to the SBF technique (Table
\ref{table:shell}). The absolute SBF magnitudes,
${\bar{M}}_{NON-SBF~DMs}$, of these shell galaxies, calculated as the
difference between the SBF apparent magnitude, $\bar{m}_{T01}$ and the
non-SBF DMs listed in Table \ref{table:shell}, are reported in
Fig. \ref{fig_shell:ps}.  The solid line in the figure represents the
empirical calibration, Eq. \ref{eq:jensen}, from
\citet[][]{Jensen+03}. Taking into account the uncertainties, the
observational data are consistent with the empirical calibration. The
median difference between the absolute SBF magnitude predicted using
Eq. \ref{eq:jensen} and ${\bar{M}}_{NON-SBF~DMs}$ is: $\langle
\bar{M}_{J03}-\bar{M}_{NON-SBF~DMs}\rangle = -0.06 \pm 0.22$ mag, so
that no systematic offset is recognized. Moreover, the data-point
nicely overlap with models. These results provide a further support in
using the SBF method to derive distances of shell ellipticals.

Relying on these agreements, and using the same set of stellar
population models matching the \citet{Jensen+03} equation, we derived:

\begin{equation}
\label{eq:calibACS1}\overline{M}_{F814W}= (-0.94 \pm 0.20)+ (2.2 \pm
0.2) \times [(F475W- F814W)_{0} - 2.0]
\end{equation}
the relation is shown in Fig. \ref{fig:fit} (panel $b$).

Following the same procedure, we derived the $\overline{M}_{I}$
versus (B-I)$_0$ calibration (see Fig. \ref{fig:fit}, panel $c$),
which extends to colors $(B-I)_{0} \leq 2.0$ the previous
calibration from \citet{Cantiello+05}. The fit to models provides:

\begin{equation}
\label{eq:calibBI} \overline{M}_{I}=\left(-1.6 \pm
0.2\right)+\left(2.1 \pm 0.2 \right)\left[(B-I)_{0}-2.0\right].
\end{equation}

Using Eq. \ref{eq:calibACS1}-\ref{eq:calibBI} and the data in
Table \ref{tab:rismag}, we obtained the DMs reported in Table
\ref{table:dm} (Col. 2 and Col. 3).

We note that the latter two relations may suffer of systematic
uncertainties which typically affect stellar population synthesis
models as, for example, the adopted library of stellar atmosphere
models, especially for cool and bright stars. In Fig. \ref{fig:fit}
(panel d) we show the changes expected if a different stellar
atmospheres library is used to obtain stellar population models in the
$F475W$ and $F814W$ filters. The new atmospheres library is the
combination of stellar models by \citet{Westera+02} for relatively hot
stars ($T_e> 4500 $ K), and by the PHOENIX models for cool stars
\citep{Brott&Hauschildt05}.  Adopting these atmosphere models, Eq.
\ref{eq:calibACS1} becomes:

$\overline{M}_{F814W}= (-0.86 \pm 0.20) + (2.3 \pm 0.2)
\times[(F475W- F814W)_{0} - 2.0]$.

The distance moduli obtained using the latter equation agree within
uncertainties with those reported in Table \ref{table:dm}.
 On average we find that the new DM varies less than 0.05
mag.

After transforming the $(B-I)_{0}$ colors in Table \ref{tab:rismag} to
$(V-I)_{0}$ color, we also derived the distance moduli of the four
galaxies using the empirical calibration, Eq. (\ref{eq:jensen}) and
the $m_{\bar{I}}$ data in the Table \ref{tab:rismag}.  To this
purpose, similarly to \citet{Cantiello+05}, we derived the (V-I)$_0$
versus (B-I)$_0$ color transformation, using the same set of models
adopted to obtain the Eqs.  \ref{eq:myjensen}-\ref{eq:calibBI}. By
fitting a straight line to the models, we derived:

$(V-I)_{0,transf}=0.47 \pm 0.02 \times (B-I)_{0}+ \left(0.21\pm
0.02\right) $

The distance moduli obtained from this procedure are in Col.  (4) of
Table \ref{table:dm}. All the DMs from the different calibrations are
in good agreement within uncertainties.

The uncertainties reported in Table \ref{table:dm} come from the
propagation of statistical uncertainties already described in
Sect. \ref{sect: dataanalysis}, and from the calibrations
uncertainty. In addition, the following systematic uncertainties
should also be taken into account: $i)$ $\sim$0.1 mag from
flat-fielding, filter zeropoint, etc., and, $ii)$ the upper limit
$\sim$0.25 mag which is the maximum uncertainty
 possible due to possible
residual shells.

Finally, we compare our distances with the ones obtained using the
Hubble law, reported in Table \ref{table:obs}. The recession velocity
adopted for each galaxy are corrected for the Virgo + Great Attractor +
Shapley Supercluster infall, based on the local velocity field model
given in \citet{Mould+00}. The good matching of our DMs with the
kinematic DMs ($\langle | DM_{SBF} - DM_{H_{0}} | < 0.2$) confirms
that it is possible to measure reliable SBF magnitudes for galaxies up
to 100 Mpc, even in the optical bandpasses, with an uncertainty
of $\pm$8 Mpc (statistical) $\pm$ 6 Mpc (systematic, $\pm$ 10 Mpc
if the upper limit uncertainty due to possible unremoved shells is
taken into account).

\subsection{H$_{0}$ determination}
\label{sect: Ho}

Since our galaxies are located beyond 4000 km/s$^{-1}$, they
constitute a good sample to determine the Hubble constant, $H_{0}$,
given that the effect of local deviation from the smooth Hubble flow
is minimized at this redshift. Using the distances based on the
calibration Eq. \ref{eq:calibACS1}, we estimate $H_{0} = 76 \pm 6$
(statistical) $\pm 5$ (systematic) [$\pm 8 $ systematic including the
upper limit uncertainty from possible unsubtracted shells] km/s/Mpc.
 It should be noted here that the $H_0$ value reported has been
derived adopting the SBF measurements and the theoretical calibrations
presented in this work, i.e. it does not suffer for the uncertainty of
the Cepheids calibration. On the other hand, the adopted calibration
suffers for the uncertainties and assumptions that are tipically
embedded in stellar population synthesis models \citep[see for
example][]{Charlot+96}. However, the reliability of the present models
is tested against known observational data (for example the empirical
calibration from \citeauthor{Jensen+03}). Such comparisons suggest
that the theoretical systematic uncertainties are not larger than the
quoted uncertainties.

Even if this determination is based on only four galaxies, it is
interesting to note that the $H_{0}$ value derived is in good
agreement with the final result from the HST Key Project Team,
$H_{0}=72\pm 4$ (statistical) $\pm 8$ (systematic) km/s/Mpc, and
with the value $H_{0}=70\pm 5$ (statistical) $\pm 6$ (systematic)
km/s/Mpc obtained by the same authors using only SBF distances
\citep{Freedman+01}. Finally, such value also agrees with the recent
values $H_{0}=73\pm4$(statistical) $\pm5$ (systematic)
km/s/Mpc determined by \citet{Riess+05}, using the multicolor light
curve shape method on two SNIa, and $H_{0}=72\pm4$ (statistical)
$\pm4$ (systematic) km/s/Mpc determined by \citet{Wang+06},
obtained by using a sample of 109 SNIa and the $\Delta C_{12}$ method.

\section{Summary and conclusion}
\label{sect: conclusions}

We presented $F814W$ SBF measurements from ACS images of four distant
shell elliptical galaxies with radial velocities between 4000 and 8000
km/s. By using the SBF method, the distance moduli of these galaxies
are derived for the first time.  We provided new calibration relations
of the absolute SBF magnitude versus the integrated color,
specifically the $\overline{M}_{F814W}$ versus
$(F475W-F814W)_{0}$. Moreover, the $\overline{M}_{I}$ versus
$(B-I)_{0}$ relation presented here extends to bluer colors the
calibration of \citet{Cantiello+05}.  The calibrations are based on
new SBF models computed with the SPoT code for the ACS and standard
filters. These models are aimed to simulate single-burst stellar
populations of age ranging from $t=1.5$ Gyr up to $t=14$ Gyr and
metallicity from Z=0.008 to Z=0.04. The use of a theoretical
calibration is important not only because it is free from the
uncertainties affecting empirical secondary distance
indicators, but also because no empirical calibration of SBF magnitudes
for these photometric bands is available in literature. To verify the
reliability of the $\overline{M}_{F814W}$ versus $(F475W-F814W)_{0}$
calibration we used exactly the same set of models to derive
SBF-vs-color relations in bands for which well established empirical
calibrations are available \citep[e.g.][]{Jensen+03,Mei+07}. As a
result, the comparison between theoretical and empirical calibrations
shows an extremely good agreement. This result added to positive tests
already performed on resolved and unresolved stellar populations
\citep[e.g.][]{Brocato+99,Brocato+00,Cantiello+03,Raimondo+05b,Fagiolini+07}
make us confident that the theoretical calibration presented here can
be safely adopted to derive distances.

On the observational side, our measurements suffer for $\sim$0.1 mag
systematic uncertainty in the SBF measurements coming from the filter
zeropoint, flat fielding, and PSF normalization. In addition, the
presence of possible unsubtracted shells can affect the estimation of
SBF amplitudes; we estimated an upper limit uncertainty of $\sim$0.25
mag from the worst case in the present data (PGC6240: the farthest
galaxy, prominent shell).

Coupling the SBF measurements with the theoretical calibrations we
find distances in agreement with the ones obtained using the Hubble
law. The present measurements enlarge the sample of galaxies beyond
100 Mpc with optical SBF distances.

Finally, using our SBF distances, we derive  $H_{0} = 76 \pm 6$
(statistical) $\pm 5$ (systematic) [$\pm 8 $ systematic when the
upper limit uncertainty from possible unsubtracted shells is included]
km/s/Mpc, in good agreement with the most recent estimations of the
Hubble constant.

\begin{acknowledgments}
It is a pleasure to acknowledge B. Patricelli for making available the
color-temperature transformation for the ACS filters computed for her
{\it Laurea} thesis.  We also thank the referee for her/his useful
suggestions which improved the paper. Financial support for this work
was provided by PRIN-INAF 2006 ``From Local to Cosmological Distances"
(P.I.: G.  Clementini).
\end{acknowledgments}




\include{tab1}
\include{tab2}
\include{tab3}
\include{tab4}
\include{tab5}


\end{document}

%% file: tab1.tex
\begin{table*}[t]
\caption{Properties of galaxies.}
\begin{center}
\tiny
\begin{tabular}{ccccccccccc}
\hline \hline
 Galaxy  &  T  & R.A. & Decl & $v_{rec}(km/s)$ & DM       &$A_{B}$ & $A_{F814W}$ & $A_{F475W}$ & F814W         & F475W \\
         &     &      &      &                 &           &        &             &             & exposure time & exposure time \\
         &     &      &      &                 &          &        &             &             & (sec)         & (sec) \\
(1 )     & (2) & (3)  & (4)  & (5)             & (6)      &  (7)   &(8)          & (9)           & (10)& (11) \\
\hline
PGC\,6510  & -3 &  26.591  & -83.400 & 4650 $\pm$ 22 & 34.0$\pm$ 0.2  & 0.588 & 0.246    & 0.492 & 2712 & 7874 \\
PGC\,10922 & -2 &  43.400  & -83.142 & 4825 $\pm$ 42 & 34.1 $\pm$ 0.2 & 0.370 & 0.156   & 0.312 & 3970 & 9350\\
PGC\,42871 & 0 & 191.022   & -34.202 & 6400 $\pm$ 36 & 34.7 $\pm$ 0.2 & 0.290 & 0.121    & 0.242 & 5908 & 8369\\
PGC\,6240  & -3 & 25.379   & -65.615 & 7936\tablenotemark{a} $\pm$ \nodata & 35.2 $\pm$ 0.5 & 0.088 & 0.036   & 0.072 & 7300 & 21440\\
\hline
\end{tabular}
\end{center}
\tablenotetext{}{ NOTE-- Col. (1): Galaxy name. Col. (2):
Morphological T-type from RC3. Cols.(3-4): Right Ascension and
Declination from RC3 (J2000). Col. (5): Recession velocity in the CMB
reference frame corrected for Virgo + Great Attractor + Shapley's
infall (data from the NED database,
www.http://nedwww.ipac.caltech.edu/). Col. (6): Kinematic Distance
Modulus obtained using $H_0=73 \pm 5$ km/s/Mpc (from NED). Col (7) B-band
extinction from \citet{Schlegel+1998}. Col. (8-9): F814W- and
F475W-band extinctions, calculated from $A_{B}$ using
\citet{Sirianni+05} coefficients. Col. (10-11): Total exposure time
for F814W and F475W images.}  \tablenotetext{a}{For this galaxy only
the Virgo infall corrected recession velocity is available.}
\label{table:obs}
\end{table*}

%% file: tab2.tex
\begin {table*}[t]
\begin{center}
\scriptsize 
\caption{Color and SBF measurements.}
\begin{tabular}{lllccccc}
\hline \hline
Galaxy & Annulus &  $P_{0}$ &  $P_{res}$ &   $\overline{\textbf{m}}_{f814W,0}$ &  (F475W-F814W)$_{0}$ & $\overline{\textbf{m}}_{I,0}$ &(B-I)$_{0}$   \\
       & (arcsec) &  (ADU/s) & (ADU/s) &  (mag)&  (mag)&  (mag) & (mag) \\
    (1)& (2) & (3) & (4) & (5) &(6) & (7) & (8) \\
\hline
 PGC\,6510  &  6-35 & 0.0031  $\pm$ 0.0001  & 0.0002 & 31.65 $\pm$ 0.04  & 1.47 $\pm$ 0.05 & 31.70 $\pm$ 0.04 &  1.86 $\pm$ 0.05  \\
 PGC\,10922 & 13-40 & 0.00225 $\pm$ 0.00004 & 0.0002 & 31.93 $\pm$ 0.04  & 1.42 $\pm$ 0.04 & 31.97 $\pm$ 0.04 &  1.79 $\pm$ 0.04 \\
 PGC\,42871 & 13-24 & 0.0045  $\pm$ 0.0002  & 0.0005 & 32.18 $\pm$ 0.04  & 1.35 $\pm$ 0.03 & 32.22 $\pm$ 0.08 &  1.70 $\pm$ 0.03 \\
 PGC\,6240  & 16-30 & 0.00125 $\pm$ 0.00006 & 0.0001 & 32.63 $\pm$ 0.06  & 1.23 $\pm$ 0.03 & 32.68 $\pm$ 0.06 &  1.57 $\pm$ 0.03 \\
    \hline
\label{tab:rismag}
\end{tabular}
\end{center}
\tablenotetext{}{NOTE-- The magnitudes are extinction and K-corrected.
  Col.  (1): Galaxy name. Col. (2): Average annulus inner-outer radii
  for SBF measurements. Col. (3) Average $P_{0}$. Col.  (4): Unmasked
  external sources fluctuation contribution $P_{res}$. Col. (5): SBF
  magnitude. Col.  (6): $(F475W-F814W)_{0}$ color. Col. (7): SBF
  magnitude in the standard photometric system. Col. (8): $(B-I)_{0}$
  integrated color.}
\end{table*}

%% file: tab3.tex
\begin{table*}[t]
\caption{SBF models.}
\begin{center}
\tiny
\begin{tabular}{ccccccccccc}
\hline \hline
 & & & & & & & & & & \\
    Age &  $\overline{F435W}$  & $\overline{ F475W}$  &  $\overline{F550M}$ & $\overline{F555W}$  & $\overline{F606W}$ & $\overline{F625W }$  & $\overline{F775W}$ & $\overline{F814W}$ &  $\overline{F850LP}$ & $(F475W-F814W)_0$\\
    (Gyr) &  (mag)    &  (mag)   & (mag)      &     (mag)&    (mag) &  (mag)  &   (mag)  & (mag)    & (mag)  & (mag)  \\
    (1) &  (2)    &  (3)   & (4)      &     (5)&    (6) &  (7)  &   (8)  & (9)    & (10)  & (11)   \\
 & & & & & & & & & &\\
\hline
\multicolumn{11}{c}{Z = 0.008} \\
\hline
   1.5 & 1.988 & 1.201 & $-$.021 &  .277   & $-$.361 & $-$.766  &   $-$2.075 &  $-$2.387 &  $-$3.090 & 1.298\\
   2.0 & 2.097 & 1.287 &  .086   &  .376   & $-$.249 & $-$.642  &  $-$1.959 &  $-$2.277 &  $-$2.987 &   1.395\\
   3.0 & 2.167 & 1.363 &  .200   &  .479   & $-$.129 & $-$.510  &  $-$1.871 &  $-$2.203 &  $-$2.928 &  1.501\\
   4.0 & 2.332 & 1.532 &  .382   &  .658   &  .055   & $-$.322  &  $-$1.708 &  $-$2.050 &  $-$2.793 & 1.504\\
   5.0 & 2.242 & 1.465 &   .353   &  .619   &  .038   & $-$.328  &  $-$1.645 &  $-$1.973 &  $-$2.696 & 1.575\\
   7.0 & 2.264 & 1.475 &.368   &  .632   &  .050   & $-$.312  &  $-$1.711 &  $-$2.061 &  $-$2.809 &  1.621\\
   9.0 & 2.429 & 1.664 &   .587   &  .843   &  .277   & $-$.079  &  $-$1.450 &  $-$1.805 &  $-$2.569 & 1.656\\
  11.0 & 2.380 & 1.636 & .587   &  .835   &  .284   & $-$.064  &  $-$1.433 &  $-$1.805 &  $-$2.601 & 1.694\\
  13.0 & 2.431 & 1.698 &  .658   &  .905   &  .354   &  .008    &  $-$1.398 &  $-$1.776 &  $-$2.579 &  1.72\\
  14.0 & 2.393 & 1.668 &   .639   &  .883   &  .337   & $-$.007  &  $-$1.412 &  $-$1.795 &  $-$2.605 & 1.737\\
\hline
\multicolumn{11}{c}{Z = 0.01} \\
\hline
   1.5 & 2.094 & 1.323 &  .113   &  .412 & $-$.230 & $-$.634 & $-$2.014 &  $-$2.347 &  $-$3.086  & 1.309\\
   2.0 & 2.270 & 1.471 &  .271   &  .563 & $-$.065 & $-$.460 & $-$1.835 &  $-$2.177 &  $-$2.935  & 1.42\\
   3.0 & 2.402 & 1.594 &  .414   &  .701 &  .080   & $-$.305 & $-$1.732 &  $-$2.094 &  $-$2.879   & 1.528\\
   4.0 & 2.489 & 1.690 &   .527   &  .810 &  .196   & $-$.185 & $-$1.657 &  $-$2.039 &  $-$2.854   & 1.573\\
   5.0 & 2.304 & 1.506 &   .368   &  .643 &  .048   & $-$.322 & $-$1.712 &  $-$2.074 &  $-$2.862   &  1.612\\
   7.0 & 2.350 & 1.572 &  .468   &  .734 &  .154   & $-$.206 & $-$1.608 &  $-$1.983 &  $-$2.783   &  1.648\\
   9.0 & 2.470 & 1.707 &   .630   &  .888 &  .325   & $-$.029 & $-$1.394 &  $-$1.771 &  $-$2.584   &  1.687\\
  11.0 & 2.465 & 1.701 &  .626   &  .883 &  .317   & $-$.036 & $-$1.427 &  $-$1.807 &  $-$2.621   &  1.725\\
  13.0 & 2.477 & 1.737 &  .690   &  .940 &  .389   &  .042   & $-$1.346 &  $-$1.739 &  $-$2.581   &  1.753\\
  14.0 & 2.572 & 1.833 &   .782   & 1.034 &  .477   &  .129   & $-$1.289 &  $-$1.694 &  $-$2.553   &  1.766\\
\hline
\multicolumn{11}{c}{Z = 0.02} \\
\hline
   1.5 & 2.579 & 1.848 & .647 &  .956 &  .311 & $-$.091 &  $-$1.582 & $-$1.994 &  $-$2.903 & 1.431\\
   2.0 & 2.723 & 1.933 &  .708 & 1.018 &  .369 & $-$.030 &  $-$1.478 & $-$1.889 &  $-$2.816 & 1.524\\
   3.0 & 2.903 & 2.096 &  .874 & 1.183 &  .539 &  .145   &  $-$1.352 & $-$1.786 &  $-$2.734 & 1.615\\
   4.0 & 2.926 & 2.139 &  .964 & 1.260 &  .642 &  .262   & $-$1.185 & $-$1.628 &  $-$2.606 & 1.673\\
   5.0 & 2.803 & 2.006 &  .832 & 1.126 &  .513 &  .137   &  $-$1.283 & $-$1.704 &  $-$2.629 & 1.703\\
   7.0 & 2.912 & 2.121 &  .974 & 1.259 &  .662 &  .294   &  $-$1.074 & $-$1.489 &  $-$2.415 &  1.749\\
   9.0 & 3.003 & 2.229 & 1.108 & 1.385 &  .800 &  .438   &  $-$.906  & $-$1.320 &  $-$2.260 & 1.793\\
  11.0 & 2.981 & 2.204 &  1.080 & 1.357 &  .769 &  .407   &  $-$.910  & $-$1.324 &  $-$2.271 & 1.826\\
  13.0 & 2.992 & 2.230 &  1.123 & 1.396 &  .814 &  .455   & $-$.901  & $-$1.337 &  $-$2.322 &  1.862\\
  14.0 & 2.979 & 2.230 &  1.147 & 1.413 &  .845 &  .493   &  $-$.820  & $-$1.248 &  $-$2.227 & 1.876\\
\hline
\multicolumn{11}{c}{Z = 0.04} \\
\hline
   1.5 & 2.903 & 2.198 &  1.035 & 1.340 &  .734 &  .350 &  $-$1.135 & $-$1.619 & $-$2.699  & 1.533\\
   2.0 & 3.057 & 2.318 & 1.146 & 1.450 &  .845 &  .465 &  $-$.999  & $-$1.482 & $-$2.564  & 1.604\\
   3.0 & 3.308 & 2.545 & 1.373 & 1.674 & 1.073 &  .696 &  $-$.744  & $-$1.248 & $-$2.390  & 1.691\\
   4.0 & 3.428 & 2.659 & 1.489 & 1.789 & 1.192 &  .819 &  $-$.614  & $-$1.121 & $-$2.277  & 1.743\\
   5.0 & 3.073 & 2.303 &  1.152 & 1.444 &  .863 &  .499 &  $-$.819  & $-$1.269 & $-$2.318  & 1.785\\
   7.0 & 3.245 & 2.453 &  1.295 & 1.587 & 1.005 &  .643 &  $-$.701  & $-$1.171 & $-$2.247  & 1.846\\
   9.0 & 3.273 & 2.494 &  1.360 & 1.645 & 1.073 &  .717 &  $-$.600  & $-$1.059 & $-$2.132  & 1.886\\
  11.0 & 3.315 & 2.552 &  1.442 & 1.719 & 1.160 &  .809 &  $-$.443  & $-$.890  & $-$1.968  & 1.922\\
  13.0 & 3.405 & 2.650 & 1.549 & 1.824 & 1.266 &  .917 &  $-$.330  & $-$.772  & $-$1.842  & 1.955\\
  14.0 & 3.281 & 2.529 &  1.438 & 1.709 & 1.157 &  .811 & $-$.449  & $-$.902  & $-$1.990  & 1.97\\
\hline
\end{tabular}
\end{center}
\tablenotetext{}{NOTE-- Col. (1): Age. Col (2-10): absolute SBF
magnitudes in various ACS photometric filters. Col(11):
(F475W-F814W)$_{0}$ integrated color.}  \label{tab:models}
\end{table*}

%% file: tab4.tex
\begin{table*}[t]
\caption{Non-SBF DMs of shell galaxies and SBF magnitudes from \citet{Tonry+01}}
\begin{center}
\small 
\begin{tabular}{ccccc}
\hline
Galaxy     & $\bar{m}_{T01}$   &       $DM$     & Distance Indicator & Reference           \\
           & (mag)       &       (mag)    &                    &                     \\
 (1)       &    (2)      &        (3)     & (4)                 &     (5)               \\
NGC\,1316  & 29.83  &31.2  $\pm$0.1 &     PNLF\tablenotemark{a}& \citet{Ferrarese+00}\\
NGC\,1344  & 29.67  &31.4  $\pm$0.2 &      PNLF                & \citet{Teodorescu+05}\\
NGC\,3923  & 30.26  &31.5  $\pm$0.2 &    GCLF                  & \citet{Sikkema+07}   \\
NGC\,4278  & 29.34  &31.1  $\pm$0.1 &     GCLF                 & \citet{Kundu+01}     \\
NGC\,5128  & 26.05  &27.7  $\pm$0.2 &   Cepheids               & \citet{Ferrarese+07} \\
NGC\,4552  & 29.39  &30.84 $\pm$0.09&   GCLF                   & \citet{Kundu+01}     \\
\hline
\label{table:shell}
\end{tabular}
\end{center}
\tablenotetext{}{NOTE-- Col. (1): Galaxy name. Col. (2): apparent SBF
magnitude measured by \citet{Tonry+01}, corrected for extinction. Col
(3): DM obtained from the distance indicator listed in Col(3). Col
(4): Reference for the DM.} \tablenotetext{a}{Planetary Nebula LF.}
\end{table*}

%% file: tab5.tex
\begin{table*}[t]
\caption{Distance Moduli.}
\begin{center}
\small 
\begin{tabular}{ccccc}
\hline
 Galaxy   &$DM_{ACS1}$\tablenotemark{a}      & $ DM_{STD,BI}$\tablenotemark{a}     & $DM_{STD,Jensen}$\tablenotemark{a} \\
 (1)      & (2)             &    (3)        &   (4)                   \\
 PGC\,6510  & 33.7 $\pm$ 0.25  & 33.6$\pm$0.25  &  33.6  $\pm$ 0.15    \\
 PGC\,10922 & 34.2 $\pm$ 0.25 & 34.1$\pm$0.25  &  34.0  $\pm$ 0.15    \\
 PGC\,42871 & 34.7 $\pm$ 0.25  & 34.6$\pm$0.25  &  34.5  $\pm$ 0.15    \\
 PGC\,6240  & 35.2 $\pm$ 0.25  & 35.2$\pm$0.25  &  35.1  $\pm$ 0.2     \\
\hline
\label{table:dm}
\end{tabular}
\end{center}
\tablenotetext{}{NOTE-- Col. (1): Galaxy name. Col. (2): DM obtained
with eq. \ref{eq:calibACS1}.  Col. (3): DM obtained with eq.
\ref{eq:calibBI}. Col(4): DM obtained with eq. \ref{eq:jensen}}
\tablenotetext{a}{Statistical uncertainties are reported. The systematic
error is $\sim 0.1$ mag (see text).}
\end{table*}

%% file: ms.bbl
\clearpage
\begin{thebibliography}{57}
\expandafter\ifx\csname natexlab\endcsname\relax\def\natexlab#1{#1}\fi

\bibitem[{{Ajhar} {et~al.}(1994){Ajhar}, {Blakeslee}, \& {Tonry}}]{Ajhar+94}
{Ajhar}, E.~A., {Blakeslee}, J.~P., \& {Tonry}, J.~L. 1994, \aj, 108, 2087

\bibitem[{{Bernstein} {et~al.}(2002){Bernstein}, {Freedman}, \&
  {Madore}}]{Bernstein+02}
{Bernstein}, R.~A., {Freedman}, W.~L., \& {Madore}, B.~F. 2002, \apj, 571, 107

\bibitem[{{Bertin} \& {Arnouts}(1996)}]{Bertin+96}
{Bertin}, E., \& {Arnouts}, S. 1996, \aaps, 117, 393

\bibitem[{{Blakeslee} {et~al.}(1999){Blakeslee}, {Ajhar}, \&
  {Tonry}}]{Blakeslee+99}
{Blakeslee}, J.~P., {Ajhar}, E.~A., \& {Tonry}, J.~L. 1999, in Astrophysics and
  Space Science Library, Vol. 237, Astrophysics and Space Science Library, ed.
  A.~{Heck} \& F.~{Caputo}, 181

\bibitem[{{Brocato} {et~al.}(2000){Brocato}, {Castellani}, {Poli}, \&
  {Raimondo}}]{Brocato+00}
{Brocato}, E., {Castellani}, V., {Poli}, F.~M., \& {Raimondo}, G. 2000, \aaps,
  146, 91

\bibitem[{{Brocato} {et~al.}(1999){Brocato}, {Castellani}, {Raimondo}, \&
  {Romaniello}}]{Brocato+99}
{Brocato}, E., {Castellani}, V., {Raimondo}, G., \& {Romaniello}, M. 1999,
  \aaps, 136, 65

\bibitem[{{Brott} \& {Hauschildt}(2005)}]{Brott&Hauschildt05}
{Brott}, I., \& {Hauschildt}, P.~H. 2005, in ESA SP-576: The Three-Dimensional
  Universe with Gaia, ed. C.~{Turon}, K.~S. {O'Flaherty}, \& M.~A.~C.
  {Perryman}, 565

\bibitem[{{Cantiello} {et~al.}(2005){Cantiello}, {Blakeslee}, {Raimondo},
  {Mei}, {Brocato}, \& {Capaccioli}}]{Cantiello+05}
{Cantiello}, M., {Blakeslee}, J.~P., {Raimondo}, G., {Mei}, S., {Brocato}, E.,
  \& {Capaccioli}, M. 2005, \apj, 634, 239

\bibitem[{{Cantiello} {et~al.}(2007){Cantiello}, {Raimondo}, {Blakeslee},
  {Brocato}, \& {Capaccioli}}]{Cantiello+07a}
{Cantiello}, M., {Raimondo}, G., {Blakeslee}, J.~P., {Brocato}, E., \&
  {Capaccioli}, M. 2007, \apj, 662, 940

\bibitem[{{Cantiello} {et~al.}(2003){Cantiello}, {Raimondo}, {Brocato}, \&
  {Capaccioli}}]{Cantiello+03}
{Cantiello}, M., {Raimondo}, G., {Brocato}, E., \& {Capaccioli}, M. 2003, \aj,
  125, 2783

\bibitem[{{Charlot} {et~al.}(1996){Charlot}, {Worthey}, \&
  {Bressan}}]{Charlot+96}
{Charlot}, S., {Worthey}, G., \& {Bressan}, A. 1996, \apj, 457, 625

\bibitem[{{Fagiolini} {et~al.}(2007){Fagiolini}, {Raimondo}, \&
  {Degl'Innocenti}}]{Fagiolini+07}
{Fagiolini}, M., {Raimondo}, G., \& {Degl'Innocenti}, S. 2007, \aap, 462, 107

\bibitem[{{Ferrarese} {et~al.}(2000){Ferrarese}, {Mould}, {Kennicutt},
  {Huchra}, {Ford}, {Freedman}, {Stetson}, {Madore}, {Sakai}, {Gibson},
  {Graham}, {Hughes}, {Illingworth}, {Kelson}, {Macri}, {Sebo}, \&
  {Silbermann}}]{Ferrarese+00}
{Ferrarese}, L. {et~al.} 2000, \apj, 529, 745

\bibitem[{{Ferrarese} {et~al.}(2007){Ferrarese}, {Mould}, {Stetson}, {Tonry},
  {Blakeslee}, \& {Ajhar}}]{Ferrarese+07}
{Ferrarese}, L., {Mould}, J.~R., {Stetson}, P.~B., {Tonry}, J.~L., {Blakeslee},
  J.~P., \& {Ajhar}, E.~A. 2007, \apj, 654, 186

\bibitem[{{Forbes} {et~al.}(1995){Forbes}, {Reitzel}, \&
  {Williger}}]{Forbes+1995}
{Forbes}, D.~A., {Reitzel}, D.~B., \& {Williger}, G.~M. 1995, \aj, 109, 1576

\bibitem[{{Freedman} {et~al.}(2001){Freedman}, {Madore}, {Gibson}, {Ferrarese},
  {Kelson}, {Sakai}, {Mould}, {Kennicutt}, {Ford}, {Graham}, {Huchra},
  {Hughes}, {Illingworth}, {Macri}, \& {Stetson}}]{Freedman+01}
{Freedman}, W.~L. {et~al.} 2001, \apj, 553, 47

\bibitem[{{Fruchter} \& {Hook}(2002)}]{Fruchter&Hook02}
{Fruchter}, A.~S., \& {Hook}, R.~N. 2002, \pasp, 114, 144

\bibitem[{{Gonz{\' a}lez} {et~al.}(2004){Gonz{\' a}lez}, {Liu}, \& {Bruzual
  A.}}]{Gonzalez+04}
{Gonz{\' a}lez}, R.~A., {Liu}, M.~C., \& {Bruzual A.}, G. 2004, \apj, 611, 270

\bibitem[{{Harris}(2001)}]{Harris+01}
{Harris}, W.~E. 2001, in Saas-Fee Advanced Course 28: Star Clusters, ed.
  L.~{Labhardt} \& B.~{Binggeli}, 223

\bibitem[{{Jedrzejewski}(1987)}]{Jedrzejewski+87}
{Jedrzejewski}, R.~I. 1987, \mnras, 226, 747

\bibitem[{{Jensen} {et~al.}(2003){Jensen}, {Tonry}, {Barris}, {Thompson},
  {Liu}, {Rieke}, {Ajhar}, \& {Blakeslee}}]{Jensen+03}
{Jensen}, J.~B., {Tonry}, J.~L., {Barris}, B.~J., {Thompson}, R.~I., {Liu},
  M.~C., {Rieke}, M.~J., {Ajhar}, E.~A., \& {Blakeslee}, J.~P. 2003, \apj, 583,
  712

\bibitem[{{Jensen} {et~al.}(2001){Jensen}, {Tonry}, {Thompson}, {Ajhar},
  {Lauer}, {Rieke}, {Postman}, \& {Liu}}]{Jensen+01}
{Jensen}, J.~B., {Tonry}, J.~L., {Thompson}, R.~I., {Ajhar}, E.~A., {Lauer},
  T.~R., {Rieke}, M.~J., {Postman}, M., \& {Liu}, M.~C. 2001, \apj, 550, 503

\bibitem[{{Jerjen} {et~al.}(2000{\natexlab{a}}){Jerjen}, {Binggeli}, \&
  {Freeman}}]{Jerjen+00a}
{Jerjen}, H., {Binggeli}, B., \& {Freeman}, K.~C. 2000{\natexlab{a}}, \aj, 119,
  593

\bibitem[{{Jerjen} {et~al.}(1998){Jerjen}, {Freeman}, \&
  {Binggeli}}]{Jerjen+98}
{Jerjen}, H., {Freeman}, K.~C., \& {Binggeli}, B. 1998, \aj, 116, 2873

\bibitem[{{Jerjen} {et~al.}(2000{\natexlab{b}}){Jerjen}, {Freeman}, \&
  {Binggeli}}]{Jerjen+00b}
---. 2000{\natexlab{b}}, \aj, 119, 166

\bibitem[{{Koekemoer} {et~al.}(2002){Koekemoer}, {Fruchter}, {Hook}, \&
  {Hack}}]{Koekemoer+02}
{Koekemoer}, A.~M., {Fruchter}, A.~S., {Hook}, R.~N., \& {Hack}, W. 2002, in
  The 2002 HST Calibration Workshop : Hubble after the Installation of the ACS
  and the NICMOS Cooling System, Proceedings of a Workshop held at the Space
  Telescope Science Institute, Baltimore, Maryland, October 17 and 18, 2002.
  Edited by Santiago Arribas, Anton Koekemoer, and Brad Whitmore. Baltimore,
  MD: Space Telescope Science Institute, 2002., p.337, ed. S.~{Arribas},
  A.~{Koekemoer}, \& B.~{Whitmore}, 337

\bibitem[{{Kundu} \& {Whitmore}(2001)}]{Kundu+01}
{Kundu}, A., \& {Whitmore}, B.~C. 2001, \aj, 121, 2950

\bibitem[{{Lee} {et~al.}(2006){Lee}, {Lee}, \& {Hwang}}]{Lee+06}
{Lee}, J.~H., {Lee}, M.~G., \& {Hwang}, H.~S. 2006, \apj, 650, 148

\bibitem[{{Malin} \& {Carter}(1983)}]{Malin&Carter83}
{Malin}, D.~F., \& {Carter}, D. 1983, \apj, 274, 534

\bibitem[{{Maybhate} {et~al.}(2007){Maybhate}, {Goudfrooij}, {Schweizer},
  {Puzia}, \& {Carter}}]{Maybhate+07}
{Maybhate}, A., {Goudfrooij}, P., {Schweizer}, F., {Puzia}, T., \& {Carter}, D.
  2007, \aj, 134, 1729

\bibitem[{{Mei} {et~al.}(2007){Mei}, {Blakeslee}, {C{\^o}t{\'e}}, {Tonry},
  {West}, {Ferrarese}, {Jord{\'a}n}, {Peng}, {Anthony}, \& {Merritt}}]{Mei+07}
{Mei}, S. {et~al.} 2007, \apj, 655, 144

\bibitem[{{Mei} {et~al.}(2005){Mei}, {Blakeslee}, {Tonry}, {Jord{\'a}n},
  {Peng}, {C{\^o}t{\'e}}, {Ferrarese}, {Merritt}, {Milosavljevi{\'c}}, \&
  {West}}]{Mei+05a}
---. 2005, \apjs, 156, 113

\bibitem[{{Mei} {et~al.}(2003){Mei}, {Scodeggio}, {Silva}, \& {Quinn}}]{Mei+03}
{Mei}, S., {Scodeggio}, M., {Silva}, D.~R., \& {Quinn}, P.~J. 2003, \aap, 399,
  441

\bibitem[{{Mieske} {et~al.}(2006){Mieske}, {Hilker}, \& {Infante}}]{Mieske+06}
{Mieske}, S., {Hilker}, M., \& {Infante}, L. 2006, \aap, 458, 1013

\bibitem[{{Mould} {et~al.}(2000){Mould}, {Huchra}, {Freedman}, {Kennicutt},
  {Ferrarese}, {Ford}, {Gibson}, {Graham}, {Hughes}, {Illingworth}, {Kelson},
  {Macri}, {Madore}, {Sakai}, {Sebo}, {Silbermann}, \& {Stetson}}]{Mould+00}
{Mould}, J.~R. {et~al.} 2000, \apj, 529, 786

\bibitem[{{Patricelli}(2006)}]{Patricelli+06}
{Patricelli}, B. 2006, Laurea thesis, Universit\'a degli studi dell'Aquila

\bibitem[{{Pence}(1986)}]{Pence86}
{Pence}, W.~D. 1986, \apj, 310, 597

\bibitem[{{Poggianti}(1997)}]{Poggianti97}
{Poggianti}, B.~M. 1997, \aaps, 122, 399

\bibitem[{{Press} {et~al.}(1992){Press}, {Teukolsky}, {Vetterling}, \&
  {Flannery}}]{Press+92}
{Press}, W.~H., {Teukolsky}, S.~A., {Vetterling}, W.~T., \& {Flannery}, B.~P.
  1992, {Numerical recipes in C. The art of scientific computing} (Cambridge:
  University Press, 1992, 2nd ed.)

\bibitem[{{Raimondo} {et~al.}(2005){Raimondo}, {Brocato}, {Cantiello}, \&
  {Capaccioli}}]{Raimondo+05b}
{Raimondo}, G., {Brocato}, E., {Cantiello}, M., \& {Capaccioli}, M. 2005, \aj,
  130, 2625

\bibitem[{{Rekola} {et~al.}(2005){Rekola}, {Jerjen}, \& {Flynn}}]{Rekola+05}
{Rekola}, R., {Jerjen}, H., \& {Flynn}, C. 2005, \aap, 437, 823

\bibitem[{{Riess} {et~al.}(2005){Riess}, {Li}, {Stetson}, {Filippenko}, {Jha},
  {Kirshner}, {Challis}, {Garnavich}, \& {Chornock}}]{Riess+05}
{Riess}, A.~G. {et~al.} 2005, \apj, 627, 579

\bibitem[{{Schlegel} {et~al.}(1998){Schlegel}, {Finkbeiner}, \&
  {Davis}}]{Schlegel+1998}
{Schlegel}, D.~J., {Finkbeiner}, D.~P., \& {Davis}, M. 1998, \apj, 500, 525

\bibitem[{{Sikkema} {et~al.}(2007){Sikkema}, {Carter}, {Peletier}, {Balcells},
  {Del Burgo}, \& {Valentijn}}]{Sikkema+07}
{Sikkema}, G., {Carter}, D., {Peletier}, R.~F., {Balcells}, M., {Del Burgo},
  C., \& {Valentijn}, E.~A. 2007, \aap, 467, 1011

\bibitem[{{Sirianni} {et~al.}(2005){Sirianni}, {Jee}, {Ben{\'{\i}}tez},
  {Blakeslee}, {Martel}, {Meurer}, {Clampin}, {De Marchi}, {Ford}, {Gilliland},
  {Hartig}, {Illingworth}, {Mack}, \& {McCann}}]{Sirianni+05}
{Sirianni}, M. {et~al.} 2005, \pasp, 117, 1049

\bibitem[{{Tamura} {et~al.}(2000){Tamura}, {Kobayashi}, {Arimoto}, {Kodama}, \&
  {Ohta}}]{Tamura+00}
{Tamura}, N., {Kobayashi}, C., {Arimoto}, N., {Kodama}, T., \& {Ohta}, K. 2000,
  \aj, 119, 2134

\bibitem[{{Teodorescu} {et~al.}(2005){Teodorescu}, {M{\'e}ndez}, {Saglia},
  {Riffeser}, {Kudritzki}, {Gerhard}, \& {Kleyna}}]{Teodorescu+05}
{Teodorescu}, A.~M., {M{\'e}ndez}, R.~H., {Saglia}, R.~P., {Riffeser}, A.,
  {Kudritzki}, R.-P., {Gerhard}, O.~E., \& {Kleyna}, J. 2005, \apj, 635, 290

\bibitem[{{Thomsen} {et~al.}(1997){Thomsen}, {Baum}, {Hammergren}, \&
  {Worthey}}]{Thomsen+97}
{Thomsen}, B., {Baum}, W.~A., {Hammergren}, M., \& {Worthey}, G. 1997, \apjl,
  483, L37

\bibitem[{{Tonry} \& {Schneider}(1988)}]{TonrySchneider88}
{Tonry}, J., \& {Schneider}, D.~P. 1988, \aj, 96, 807

\bibitem[{{Tonry} {et~al.}(1990){Tonry}, {Ajhar}, \& {Luppino}}]{Tonry+90}
{Tonry}, J.~L., {Ajhar}, E.~A., \& {Luppino}, G.~A. 1990, \aj, 100, 1416

\bibitem[{{Tonry} {et~al.}(1997){Tonry}, {Blakeslee}, {Ajhar}, \&
  {Dressler}}]{Tonry+97}
{Tonry}, J.~L., {Blakeslee}, J.~P., {Ajhar}, E.~A., \& {Dressler}, A. 1997,
  \apj, 475, 399

\bibitem[{{Tonry} {et~al.}(2001){Tonry}, {Dressler}, {Blakeslee}, {Ajhar},
  {Fletcher}, {Luppino}, {Metzger}, \& {Moore}}]{Tonry+01}
{Tonry}, J.~L., {Dressler}, A., {Blakeslee}, J.~P., {Ajhar}, E.~A., {Fletcher},
  A.~B., {Luppino}, G.~A., {Metzger}, M.~R., \& {Moore}, C.~B. 2001, \apj, 546,
  681

\bibitem[{{Turnbull} {et~al.}(1999){Turnbull}, {Bridges}, \&
  {Carter}}]{Turnbull+99}
{Turnbull}, A.~J., {Bridges}, T.~J., \& {Carter}, D. 1999, \mnras, 307, 967

\bibitem[{{Udalski} {et~al.}(1999){Udalski}, {Soszynski}, {Szymanski},
  {Kubiak}, {Pietrzynski}, {Wozniak}, \& {Zebrun}}]{Udalsky+99}
{Udalski}, A., {Soszynski}, I., {Szymanski}, M., {Kubiak}, M., {Pietrzynski},
  G., {Wozniak}, P., \& {Zebrun}, K. 1999, Acta Astronomica, 49, 223

\bibitem[{{Wang} {et~al.}(2006){Wang}, {Wang}, {Pain}, {Zhou}, \&
  {Li}}]{Wang+06}
{Wang}, X., {Wang}, L., {Pain}, R., {Zhou}, X., \& {Li}, Z. 2006, \apj, 645,
  488

\bibitem[{{Westera} {et~al.}(2002){Westera}, {Lejeune}, {Buser}, {Cuisinier},
  \& {Bruzual}}]{Westera+02}
{Westera}, P., {Lejeune}, T., {Buser}, R., {Cuisinier}, F., \& {Bruzual}, G.
  2002, \aap, 381, 524

\bibitem[{{Wilkinson} {et~al.}(2000){Wilkinson}, {Prieur}, {Lemoine}, {Carter},
  {Malin}, \& {Sparks}}]{Wilkinson+00}
{Wilkinson}, A., {Prieur}, J.-L., {Lemoine}, R., {Carter}, D., {Malin}, D., \&
  {Sparks}, W.~B. 2000, \mnras, 319, 977

\end{thebibliography}
